\documentclass[%
reprint,
superscriptaddress,
preprintnumbers,
nofootinbib,
nobibnotes,
amsmath,amssymb,
aps,
prc,
]{revtex4-2}
\usepackage[pdftex]{graphicx}
\usepackage{dcolumn}
\usepackage{bm}
\usepackage[pdftex]{color}
\usepackage{CJK}
\usepackage[T1]{fontenc}
\usepackage{mathrsfs}
\def\ve#1{{\bm{#1}}}
\def\nuc#1#2#3{{}^{#2}_{#3}\mathrm{#1}}
\def\urm#1{\scriptstyle{\text{\textrm{\textmd{\textup{#1}}}}}}

\let\temp\epsilon
\let\epsilon\varepsilon
\let\varepsilon\temp
\let\temp\relax
\let\temp\phi
\let\phi\varphi
\let\varphi\temp
\let\temp\relax
\begin{document}
% 
%%%%%%%%%%%%%%%%%%%%%%%%%%%%%%%%%%%%%%%%%%%%%%%%%% 
\begin{CJK*}{UTF8}{}
  \preprint{RIKEN-iTHEMS-Report-25}
  \title{Theoretical evaluation of decay mode of $\nuc{Th}{229m}{}$ in solid samples}
  \author{Ryotaro Masuda (\CJKfamily{min}{益田遼太郎})}
  \email{
    masudar21@chem.sci.osaka-u.ac.jp}
  \affiliation{
    Department of Chemistry, Graduate School of Science, The University of Osaka,
    Toyonaka 560-0043, Japan}
  \author{Tomoya Naito (\CJKfamily{min}{内藤智也})}
  \email{
    tnaito@ribf.riken.jp}
  \affiliation{
    RIKEN Center for Interdisciplinary Theoretical and Mathematical Sciences (iTHEMS),
    Wako 351-0198, Japan}
  \affiliation{
    Department of Physics, Graduate School of Science, The University of Tokyo,
    Tokyo 113-0033, Japan}
  \affiliation{
    Department of Nuclear Engineering and Management, Graduate School of Engineering, The University of Tokyo,
    Tokyo 113-8656, Japan}
  \author{Masashi Kaneko (\CJKfamily{min}{金子政志})}
  \affiliation{
    Department of Chemistry, Graduate School of Science, The University of Osaka,
    Toyonaka 560-0043, Japan}
  \author{Hiroyuki Kazama (\CJKfamily{min}{風間裕行})}
  \affiliation{
    Department of Chemistry, Graduate School of Science, The University of Osaka,
    Toyonaka 560-0043, Japan}
  \author{So Hashiba (\CJKfamily{min}{橋場奏})}
  \affiliation{
    Department of Chemistry, Graduate School of Science, The University of Osaka,
    Toyonaka 560-0043, Japan}
  \author{Kosuke Misawa (\CJKfamily{min}{三澤宏介})}
  \affiliation{
    Department of Chemistry, Graduate School of Science, The University of Osaka,
    Toyonaka 560-0043, Japan}
  \author{Yoshitaka Kasamatsu (\CJKfamily{min}{笠松良崇})}
  \affiliation{
    Department of Chemistry, Graduate School of Science, The University of Osaka,
    Toyonaka 560-0043, Japan}
  \date{\today}
  %%%%%%%%%%%%%%%%%%%%%%%%%%%%%%%%%%%%%%%%%%%%%%%%%% 
  \begin{abstract}
    The excitation energy of $ \nuc{Th}{229m}{} $ is extremely low at $ 8.4 \, \mathrm{eV} $; thus, this isotope exhibits changes in its decay modes depending on the chemical state, specifically the outermost electronic states.
    However, the reported half-lives of the $ \gamma $-ray transition are not consistent among the previous experiments.
    In this study, we investigate the chemical states of $ \nuc{Th}{229m}{} $ by density functional theory calculations.
    Based on these results, we evaluate the relationship between the experimental half-life of each sample and the electronic state of $ \mathrm{Th} $.
    The calculation results indicate that ion trap method, $ \mathrm{Ca} \mathrm{F}_2 $ model and $ \mathrm{Mg} \mathrm{F}_2 $ one decay only via the $ \gamma $-ray transition,
    whereas $ \mathrm{Li} \mathrm{Sr} \mathrm{Al} \mathrm{F}_6 $ one decays via the $ \gamma $-ray transition and has a possibility of decay via internal conversion and electron bridge.
  \end{abstract}
  \maketitle
\end{CJK*}
%%%%%%%%%%%%%%%%%%%%%%%%%%%%%%%%%%%%%%%%%%%%%%%%%% 
% 
\section{Introduction}
\par
The decay constant, accordingly the half-life, of an atomic nucleus is generally considered independent of its electronic structure such as chemical states.
This is because the spatial extent and energy of the nucleus differ from those of orbital electrons by approximately five orders of the magnitude, allowing them to be treated independently in most cases. 
Nevertheless, nuclear decay modes involving interactions between the nucleus and orbital electrons do exist, for instance, internal conversion (IC) and electron capture (EC), while we do not discuss EC process in this paper.
IC process is a de-excitation where an excited nucleus transfers its energy to an orbital electron, which is emitted as an IC electron. 
Since only inner-shell electrons, particularly \textit{K}-shell electrons, participate in an IC process, chemical states hardly affect an IC property, leaving the decay constant unchanged.
However, there are some nuclei with low excitation energies  ($ E_{\urm{IS}} $) that can interact with outer-shell electrons and chemical states may change its decay constant.
For example, $ \nuc{U}{235m}{} $ has $ E_{\urm{IS}} $ of $ 76.7 \, \mathrm{eV} $~\cite{
  PhysRevC.97.054310},
which decays only via IC process, and up to a $ 10 \, \% $ change in half-life has been observed depending on chemical states~\cite{
  Mevergnies1972Phys.Rev.Lett.29_1188,
  Mevergnies1974Phys.Lett.B49_428}.
\par
There is another example, that is, $ \nuc{Th}{229m}{} $, the first excited nuclear state of $ \nuc{Th}{229}{} $.
The $ E_{\urm{IS}} $ of $ \nuc{Th}{229m}{} $ is reported to be approximately $ 8.4 \, \mathrm{eV} $
(wavelength $ \simeq 150 \, \mathrm{nm} $)~\cite{
  Kremar2023Nature.617_706,
  Yamaguchi2024Nature.629_62,
  Tiedau2024Phys.Rev.Lett.132_182501,
  Elwell2024Phys.Rev.Lett.133_013201,
  Hiraki2024Nat.Commun.15_5536},
comparable with the energy scale of chemical bonds.
Due to its lowest $ E_{\urm{IS}} $, $ \nuc{Th}{229m}{} $ has been attracted attention for two main reasons.
\par
First, the decay mode of $ \nuc{Th}{229m}{} $ can vary depending on its chemical state due to the extremely low $ E_{\urm{IS}} $. 
The possible decay modes are known as IC, $ \gamma $-ray transition, and electron bridge (EB)~\cite{
  Strizhov1991,
  PhysRevA.111.L041103,
  Karpeshin2018Nucl.Phys.A969_173-183},
while EB process has not been reported yet.
EB process involves nuclear de-excitation coupled with electronic excitation.
The ionization potential of the neutral $ \mathrm{Th} $ and several $ \mathrm{Th} $ ions are summarized in Table~\ref{tab:IE}.
There are no data of $ \mathrm{Th}^{4+} $,
because it has the same electron configuration as $ \mathrm{Rn} $ and thus it is rather stable.
The ionization potential of the neutral $\mathrm{Th}$ atom, the binding energy of the outermost electron (HOMO), is $ 6.3 \, \mathrm{eV} $~\cite{
  VonDerWenseTheEuropeanPhysicalJournalA56_277},
which is lower than $ E_{\urm{IS}} $ ($ 8.4 \, \mathrm{eV} $); thus, $ \nuc{Th}{229m}{} $ can de-excite via IC. 
Indeed, IC electron from $ \nuc{Th}{229m}{}^{2+} $ has been observed using the ion trap method~\cite{
  VonDerWenseNature553_042501} 
and the IC half-life of $ \nuc{Th}{229m}{} $ on the metallic surface of $ \mathrm{Ni} $ is reported to be $ 7 \, \mathrm{\mu s} $~\cite{
  PhysRevLett.118.042501}.
In contrast, the binding energy of orbital electrons is larger than $ E_{\urm{IS}} $ for $ \mathrm{Th} $ ions.
Hence, IC energy is not large enough to emit an outermost electron and therefore IC does not occur. 
Consequently, $ \nuc{Th}{229m}{} $ ions can decay via the $ \gamma $-ray transition or EB.
\begin{table}[tb]
  \centering
  \caption{
    Ionization potential of $ \mathrm{Th}^{n+} $ ($ n = 0 $--$ 3 $)~\cite{
      VonDerWenseTheEuropeanPhysicalJournalA56_277}.}
  \label{tab:IE}
  \begin{ruledtabular}
    \begin{tabular}{ld}
      \multicolumn{1}{l}{State of $ \mathrm{Th} $} & \multicolumn{1}{c}{Ionization potential ($ \mathrm{eV} $)} \\
      \hline
      $ \mathrm{Th} $      &  6.3 \\
      $ \mathrm{Th}^{+} $  & 12.1 \\
      $ \mathrm{Th}^{2+} $ & 18.3 \\
      $ \mathrm{Th}^{3+} $ & 28.8 \\
    \end{tabular}
  \end{ruledtabular}
\end{table}
\begin{table*}[tb]
  \centering
  \caption{
    Observed $ \gamma $ decay of $ \nuc{Th}{229m}{} $ in different samples.
    The excitation energy $ E_{\urm{IS}} $, the half-life of $ \nuc{Th}{229m}{} $ in vacuum, and that in crystal are listed,
    where no reported case is indicated as ``---''.
    The previous work of $ \mathrm{Li} \mathrm{Sr} \mathrm{Al} \mathrm{F}_6 $ reported its lifetime~\cite{
      Elwell2024Phys.Rev.Lett.133_013201},
    which is converted to the half-life here.}
  \label{tab:gamma}
  \begin{ruledtabular}
    \begin{tabular}{ldddc}
      \multicolumn{1}{l}{Sample} & \multicolumn{1}{c}{$ E_{\urm{IS}} $ ($ \mathrm{eV} $)} & \multicolumn{2}{c}{Half-life ($ \mathrm{s} $)} & \multicolumn{1}{c}{Ref.} \\
                                 & & \multicolumn{1}{c}{In vacuum} & \multicolumn{1}{c}{In crystal} & \\
      \hline
      Ion trap
                                 & \multicolumn{1}{c}{---} & 1400_{-300}^{+600} & \multicolumn{1}{c}{---} & \cite{Yamaguchi2024Nature.629_62} \\
      $ \mathrm{Ca} \mathrm{F}_2 $
                                 & 8.35574 & 1740(50) & 630(15) & \cite{Tiedau2024Phys.Rev.Lett.132_182501} \\
      $ \mathrm{Ca} \mathrm{F}_2 $
                                 & 8.367(24) & 1790 (64)_{\urm{stat}} (80)_{\urm{sys}} & 447(25) & \cite{Hiraki2024Nat.Commun.15_5536} \\
      $ \mathrm{Mg} \mathrm{F}_2 $
                                 & 8.338(24) & 2210(304) & 670(102) & \cite{Kremar2023Nature.617_706} \\
      $ \mathrm{Li} \mathrm{Sr} \mathrm{Al} \mathrm{F}_6 $
                                 & 8.355733 (2)_{\urm{stat}} (10)_{\urm{sys}} & 1289(30)_{\urm{stat}}(45)_{\urm{sys}} & 394(9)_{\urm{stat}}(14)_{\urm{sys}} & \cite{Elwell2024Phys.Rev.Lett.133_013201} \\
    \end{tabular}
  \end{ruledtabular}
\end{table*}
Second, it enables the implementation of a nuclear clock through the $ \gamma $-ray transition of $ \nuc{Th}{229m}{} $~\cite{
  Peik2003Europhys.Lett.61_181,
  PhysRevA.111.053109}.
Such a nuclear clock is expected to contribute significantly to society, enabling applications such as earthquake prediction~\cite{
  Peik_2021}
and resource exploration~\cite{
  Flambaum2006PhysicalReviewLetters.97_092502}.
\par
We summarize the reported $ E_{\urm{IS}} $ and $ \gamma $-ray half-lives of $ \nuc{Th}{229m}{} $ in Table~\ref{tab:gamma}.
In ion trap experiment, $ \nuc{Th}{229m}{}^{3+} $ ions,
expected to decay via the $ \gamma $-ray transition in a vacuum,
were selectively trapped and observed with a reported half-life of $ 1400_{-300}^{+600} \, \mathrm{s} $~\cite{
  Yamaguchi2024Nature.629_62}.
Additionally, $ \gamma $ ray has been observed from $ \nuc{Th}{229m}{} $ doped into crystals,
such as $ \mathrm{Ca} \mathrm{F}_2 $,
$ \mathrm{Mg} \mathrm{F}_2 $,
and $ \mathrm{Li} \mathrm{Sr} \mathrm{Al} \mathrm{F}_6 $~\cite{
  Kremar2023Nature.617_706,
  Tiedau2024Phys.Rev.Lett.132_182501,
  Elwell2024Phys.Rev.Lett.133_013201,
  Hiraki2024Nat.Commun.15_5536}.
The half-lives measured in crystals were consistently shorter than the ion trap method
(``In crystal'' in Table~\ref{tab:gamma}).
It is indispensable to convert the half-life in crystal to those in vacuum by multiplying by the cube of the refractive index, whose values are shown as ``In vacuum'' in Table~\ref{tab:gamma}~\cite{
  TkalyaJetpLett.71_311-313}.
These values are inconsistent with each other, while all of them are consistent with ion trap method.
One possible reason is that the refractive-index corrections are inaccurate~\cite{
  TkalyaJetpLett.71_311-313}
since doping $ \mathrm{Th} $ into crystals may change the refractive index near the $ \mathrm{Th} $ atoms locally. 
The other possible reason is that the outer-shell electronic structure of $ \mathrm{Th} $ in a crystal could be different and thus change the decay modes of $ \nuc{Th}{229m}{} $.
\par
The theoretical calculation of the decay and half-life of $ \nuc{Th}{229m}{} $
based on the density functional theory (DFT) and configuration interaction method had been reported~\cite{
  Karpeshin2018Nucl.Phys.A969_173-183,
  MULLER201784,
  PORSEV2021,
  PhysRevLett.105.182501}.
However, these calculations had focused on $ \mathrm{Th} $ in an isolated atomic state, with emphasizing nuclear properties.
Recently, there has been a report that performed DFT calculations on $ \nuc{Th}{229m}{} $-doped $ \mathrm{Ca} \mathrm{F}_2 $ and $ \mathrm{Li} \mathrm{Sr} \mathrm{Al} \mathrm{F}_6 $ crystals, analyzed the density of states, and carried out a quantitative discussion on IC~\cite{
  9s8f-hv1f},
while they did not perform the systematic study on the decay modes.
\par
In this study, we calculated the electronic structure of the $ \mathrm{Th} $  atom that mimics the condition of ion trap method and $ \mathrm{Th} $ doped in crystal. 
We selected $ \mathrm{Ca} \mathrm{F}_2 $,
$ \mathrm{Mg} \mathrm{F}_2 $, and
$ \mathrm{Li} \mathrm{Sr} \mathrm{Al} \mathrm{F}_6 $
as examples.
To investigate the electronic structure around $ \mathrm{Th} $,
consisting of $ \mathrm{Th} $ and its neighboring atoms were extracted from optimized structure,
and the natural orbital analysis was performed with the relativistic correlation.
By focusing on changes in the electronic state of $ \mathrm{Th} $, we explored possible reasons for variations in the half-life.

\section{Identification of crystal structure}
\label{2}
\par
We calculated crystals
($ \mathrm{Ca}\mathrm{F}_2 $, $ \mathrm{Mg}\mathrm{F}_2 $,
and $ \mathrm{Li} \mathrm{Sr} \mathrm{Al} \mathrm{F}_6 $) in which a $ \mathrm{Th} $ atom was doped.
We considered two models of the placement of $ \mathrm{Th} $.
One model was a $ \mathrm{Th} $ atom in the gap space of the crystals.
The other was a replacing of a $ \mathrm{Ca}^{2+} $, $ \mathrm{Mg}^{2+} $, $ \mathrm{Al}^{3+} $, or $ \mathrm{Sr}^{2+} $ ion with $ \mathrm{Th} $.
Hereinafter, the former and latter models are, respectively, abbreviated as
``$ \mathrm{X} $ gap''
and
``$ \mathrm{X} $-$ \mathrm{Y} $ replaced,''
where
$ X $ is $ \mathrm{Ca}\mathrm{F}_2 $, $ \mathrm{Mg}\mathrm{F}_2 $, or
$ \mathrm{Li} \mathrm{Sr} \mathrm{Al} \mathrm{F}_6 $,
and
$ Y $ is $ \mathrm{Ca} $, $ \mathrm{Mg} $, $ \mathrm{Sr} $, or $ \mathrm{Al} $.
In both cases, the doped $ \mathrm{Th} $ causes structural distortion, which was considered in structural optimization.
The initial charges of $ \mathrm{Th} $ were considered from $ 0 $ to $ +4 $
and structural optimizations were performed for each initial charge.
These models were calculated using a $ 3 \times 3 \times 3 $ supercell with $ 4 \times 4 \times 4 $ $ \ve{K} $ points.
These calculations were preformed using the periodic boundary condition.
\begin{figure}
  \centering
  \includegraphics[width=1.0\linewidth]{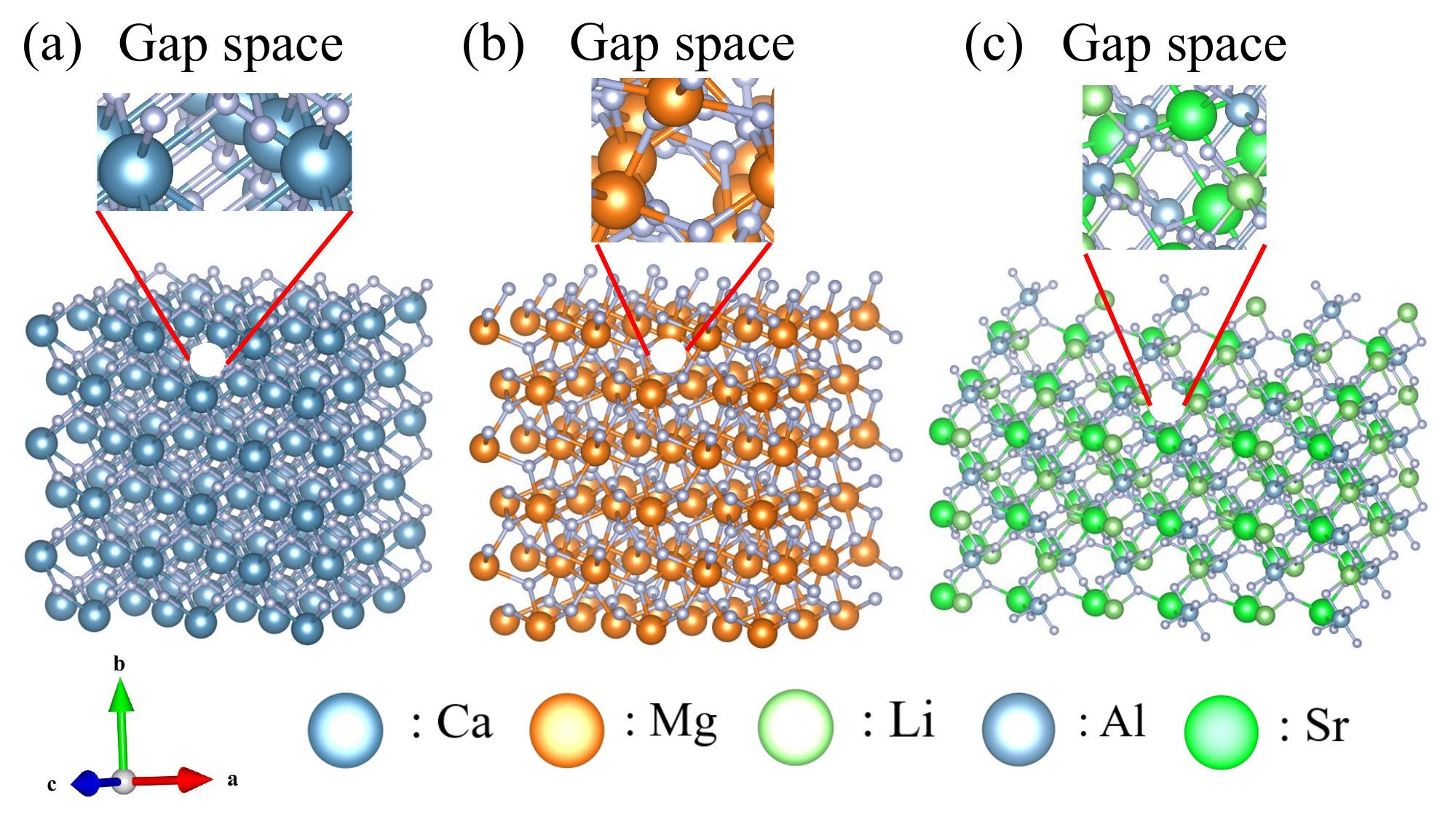}
  \caption{
    Crystal structure with $ 3 \times 3 \times 3 $ supercell model
    of 
    (a) $ \mathrm{Ca}\mathrm{F}_2 $,
    (b) $ \mathrm{Mg}\mathrm{F}_2 $ and,
    (c) $ \mathrm{Li} \mathrm{Sr} \mathrm{Al} \mathrm{F}_6 $.
    The white-filled sphericals indicate the gap spaces.}
  \label{QE-model}
\end{figure}
\par
The initial crystal structures of
$ \mathrm{Ca} \mathrm{F}_2 $,
$ \mathrm{Mg} \mathrm{F}_2 $,
and $ \mathrm{Li} \mathrm{Sr} \mathrm{Al} \mathrm{F}_6 $
were assumed as the space group of
$ \mathit{F} \mathit{m} \overline{3} \mathit{m} $ (225)~\cite{
  osti_1201632},
$ \mathit{P} \mathit{a} \overline{3} $ (205)~\cite{
  osti_1192516},
and 
$ \mathit{P} \overline{3} 1 \mathit{c} $ (163)~\cite{
  osti_1281377},
respectively.
Their crystal structures are shown in Fig.~\ref{QE-model}.
The lattice constants were optimized through DFT~\cite{
  Hohenberg1964Phys.Rev.136_B864,
  Kohn1965Phys.Rev.140_A1133,
  Kohn1999Rev.Mod.Phys.71_1253}
using the open-source code \textsc{quantum espresso} ver.~7.0.3~\cite{
  Giannozzi_2009}
with employing the PBE exchange-correlation functional~\cite{
  Perdew1996Phys.Rev.B54_16533,
  Perdew1996Phys.Rev.Lett.77_3865}
with projector augmented wave pseudopotentials~\cite{TORRENT20101862}
generated by \textsc{pslibrary}~\cite{
  DalCorso2014Comput.Mater.Sci.95_337}.
\par
The lattice constants obtained without embedding a $ \mathrm{Th} $ atom are summarized in Table~\ref{tab:lattice_const}.
Tables~\ref{tab:distance_gap} and \ref{tab:distance_replace} summarize the bond lengths
around the $ \mathrm{Th} $ atom obtained by the structural optimization,
which would be used for a further calculation shown in Sec.~\ref{sec:ADF}.
For all the cases, 
regardless of whether the $ \mathrm{Th} $ atom was in a gap or replaced site,
the nearest neighbor atoms to the $ \mathrm{Th} $ atom were a $ \mathrm{F} $ atom. 
The distance between the $ \mathrm{Th} $ atom and the $ \mathrm{F} $ one  varied depending on the initial charge of $ \mathrm{Th} $. 
This trend indicates that the distance decreases as the charge of the $ \mathrm{Th} $ atom increases, suggesting that the $ \mathrm{Th} $-$ \mathrm{F} $ interaction becomes stronger with increasing the $ \mathrm{Th} $ charge.
\begin{table}[tb]
  \centering
  \caption{Optimal lattice constant of each crystal obtained by DFT.
    The detail is shown in the text.}
  \label{tab:lattice_const}
  \begin{ruledtabular}
    \begin{tabular}{lddd}
      \multicolumn{1}{l}{Crystal} & \multicolumn{1}{c}{$ \mathrm{a} $ (\AA)} & \multicolumn{1}{c}{$ \mathrm{b} $ (\AA)} & \multicolumn{1}{c}{$ \mathrm{c} $ (\AA)} \\
      \hline
      $ \mathrm{Ca} \mathrm{F}_2 $                         & 4.72 & 4.72 &  4.72 \\
      $ \mathrm{Mg} \mathrm{F}_2 $                         & 4.34 & 4.34 &  2.91 \\
      $ \mathrm{Li} \mathrm{Sr} \mathrm{Al} \mathrm{F}_6 $ & 5.08 & 5.08 & 10.21 \\
    \end{tabular}
  \end{ruledtabular}
\end{table}
\begin{table*}[tb]
  \centering
  \caption{Distance between the $ \mathrm{Th} $ atom and the nearest metallic or a $ \mathrm{F} $ atom in the gap model.
    All the results are shown in \AA.
    The detail can be found in the text.}
  \label{tab:distance_gap}
  \begin{ruledtabular}
    \begin{tabular}{llddddd}
      \multicolumn{1}{l}{Model} & \multicolumn{1}{l}{Bond} & \multicolumn{1}{c}{$ \mathrm{Th} $} & \multicolumn{1}{c}{$ \mathrm{Th}^{+} $} & \multicolumn{1}{c}{$ \mathrm{Th}^{2+} $} & \multicolumn{1}{c}{$ \mathrm{Th}^{3+} $} & \multicolumn{1}{c}{$ \mathrm{Th}^{4+} $} \\
      \hline
      $ \mathrm{Ca} \mathrm{F}_2 $-gap
                                & $ \mathrm{Th} $-$ \mathrm{Ca} $ & 2.99 & 3.03 & 2.99 & 3.06 & 3.05 \\
                                & $ \mathrm{Th} $-$ \mathrm{F} $  & 2.48 & 2.44 & 2.48 & 2.35 & 2.36 \\
      $ \mathrm{Mg} \mathrm{F}_2 $-gap
                                & $ \mathrm{Th} $-$ \mathrm{Mg} $ & 2.51 & 2.51 & 2.51 & 2.51 & 2.51 \\
                                & $ \mathrm{Th} $-$ \mathrm{F} $  & 2.34 & 2.31 & 2.26 & 2.22 & 2.22 \\
      $ \mathrm{Li} \mathrm{Sr} \mathrm{Al} \mathrm{F}_6 $-gap
                                & $ \mathrm{Th} $-$ \mathrm{Sr} $ & 2.91 & 2.99 & 2.94 & 2.76 & 2.87 \\
                                & $ \mathrm{Th} $-$ \mathrm{Al} $ & 3.16 & 3.13 & 3.15 & 2.94 & 3.07 \\
                                & $ \mathrm{Th} $-$ \mathrm{Li} $ & 2.87 & 3.00 & 2.95 & 2.94 & 2.94 \\
                                & $ \mathrm{Th} $-$ \mathrm{F} $  & 2.27 & 2.29 & 2.27 & 2.25 & 2.24 \\
    \end{tabular}
  \end{ruledtabular}
\end{table*}
\begin{table*}[tb]
  \centering
  \caption{Same as Table~\ref{tab:distance_gap} but for the replaced model.}
  \label{tab:distance_replace}
  \begin{ruledtabular}
    \begin{tabular}{llddddd}
      \multicolumn{1}{l}{Model} & \multicolumn{1}{l}{Bond} & \multicolumn{1}{c}{$ \mathrm{Th} $} & \multicolumn{1}{c}{$ \mathrm{Th}^{+} $} & \multicolumn{1}{c}{$ \mathrm{Th}^{2+} $} & \multicolumn{1}{c}{$ \mathrm{Th}^{3+} $} & \multicolumn{1}{c}{$ \mathrm{Th}^{4+} $} \\
      \hline
      $ \mathrm{Ca} \mathrm{F}_2 $-$ \mathrm{Ca} $ replaced
                                & $ \mathrm{Th} $-$ \mathrm{Ca} $ & 3.90 & 3.90 & 3.90 & 3.90 & 3.90 \\
                                & $ \mathrm{Th} $-$ \mathrm{F} $  & 2.63 & 2.55 & 2.51 & 2.49 & 2.40 \\
      $ \mathrm{Mg} \mathrm{F}_2 $-$ \mathrm{Mg} $ replaced
                                & $ \mathrm{Th} $-$ \mathrm{Mg} $ & 2.51 & 2.51 & 2.51 & 2.51 & 2.51 \\
                                & $ \mathrm{Th} $-$ \mathrm{F} $  & 2.49 & 2.43 & 2.37 & 2.31 & 2.30 \\
      $ \mathrm{Li} \mathrm{Sr} \mathrm{Al} \mathrm{F}_6 $-$ \mathrm{Al} $ replaced
                                & $ \mathrm{Th} $-$ \mathrm{Sr} $ & 4.00 & 4.08 & 4.08 & 4.08 & 4.08 \\
                                & $ \mathrm{Th} $-$ \mathrm{Al} $ & 5.85 & 5.88 & 5.88 & 5.88 & 5.88 \\
                                & $ \mathrm{Th} $-$ \mathrm{Li} $ & 3.20 & 3.20 & 3.20 & 3.20 & 3.19 \\
                                & $ \mathrm{Th} $-$ \mathrm{F} $  & 2.27 & 2.22 & 2.22 & 2.22 & 2.28 \\
      $ \mathrm{Li} \mathrm{Sr} \mathrm{Al} \mathrm{F}_6 $-$ \mathrm{Sr} $ replaced
                                & $ \mathrm{Th} $-$ \mathrm{Sr} $ & 5.15 & 5.19 & 5.20 & 5.21 & 5.14 \\
                                & $ \mathrm{Th} $-$ \mathrm{Al} $ & 3.99 & 3.91 & 3.91 & 3.98 & 3.93 \\
                                & $ \mathrm{Th} $-$ \mathrm{Li} $ & 3.99 & 3.97 & 3.98 & 3.95 & 3.95 \\
                                & $ \mathrm{Th} $-$ \mathrm{F} $  & 2.41 & 2.26 & 2.27 & 2.27 & 2.26 \\
    \end{tabular}
  \end{ruledtabular}
\end{table*}
\section{Band analysis}
\label{sec:ADF}
\subsection{Calculation setup}
\par
\begin{figure}[tb]
  \centering
  \includegraphics[width=1.0\linewidth]{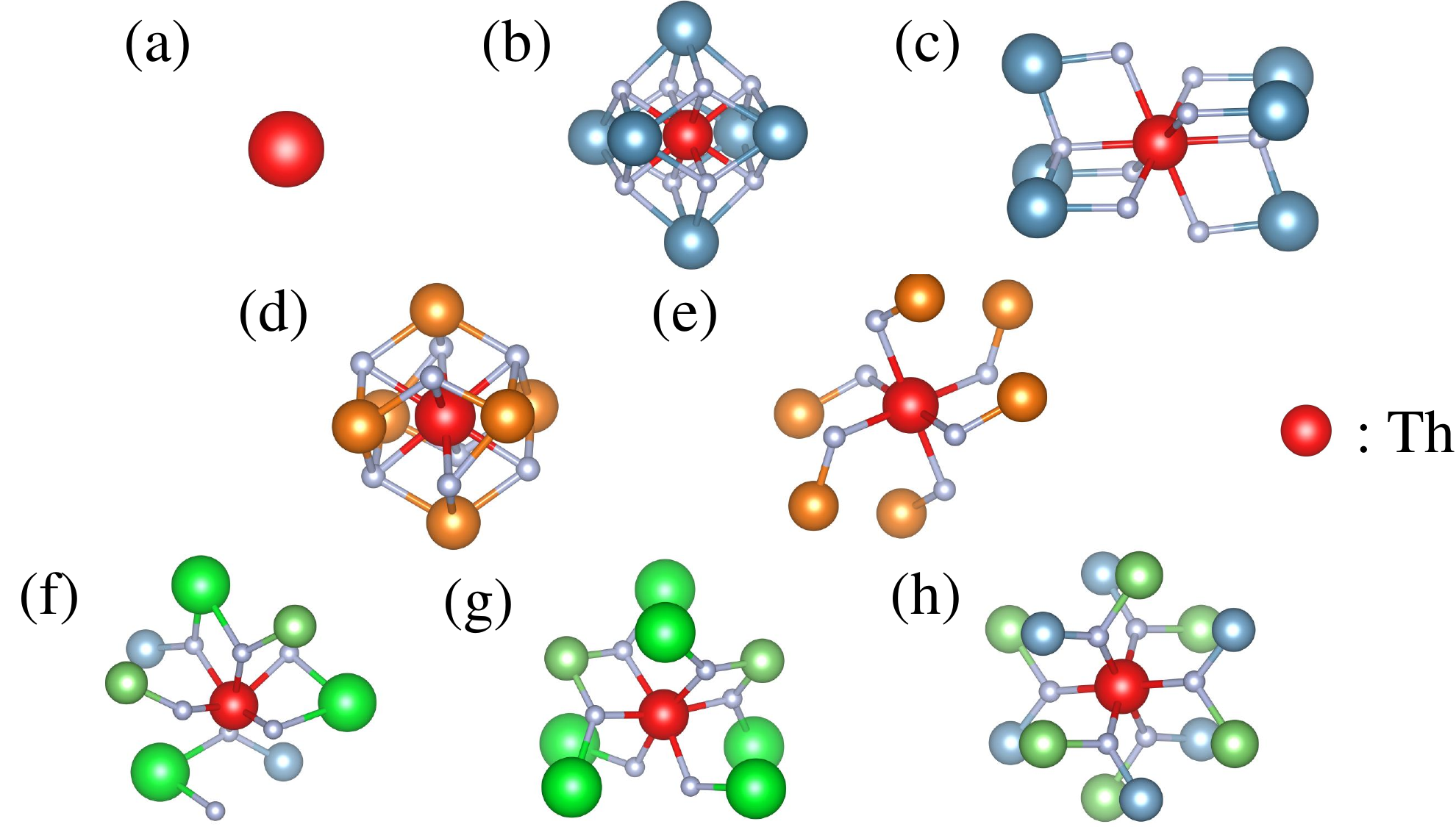}
  \caption{Considered structure for 
    (a) a $ \mathrm{Th} $ atom simulating ion trap method;
    (b) $ \mathrm{Ca} \mathrm{F}_2 $ gap model;
    (c) $ \mathrm{Ca} \mathrm{F}_2 $-$ \mathrm{Ca} $ replaced model;
    (d) $ \mathrm{Mg} \mathrm{F}_2 $ gap model;
    (e) $ \mathrm{Mg} \mathrm{F}_2 $-$ \mathrm{Mg} $ replaced model;
    (f) $ \mathrm{Li} \mathrm{Sr} \mathrm{Al} \mathrm{F}_6 $ gap model;
    (g) $ \mathrm{Li} \mathrm{Sr} \mathrm{Al} \mathrm{F}_6 $-$ \mathrm{Al} $ replaced model; 
    (h) $ \mathrm{Li} \mathrm{Sr} \mathrm{Al} \mathrm{F}_6 $-$ \mathrm{Sr} $ replaced model.
    The same structures were adopted for all the initial charges of $ \mathrm{Th} $.
    Except for $ \mathrm{Th} $, the colors are the same as shown in Fig.~\ref{QE-model}.}
  \label{ADF-model}
\end{figure}

To investigate the electronic structure of the $ \mathrm{Th} $ atom embedded in a crystal,
the code named \textsc{amsterdam density functional} (ADF) version~2023.1~\cite{https://doi.org/10.1002/jcc.1056}
was used with employing the QZ4P~\cite{https://doi.org/10.1002/jcc.10255} 
basis set and the B3LYP exchange-correlation functional~\cite{10.1063/1.464913}
without a frozen core.
The zeroth-order relativistic approximation (ZORA)~\cite{10.1063/1.466059}
was used to consider the relativistic effect to obtain the binding energy of stabilized charge of $ \mathrm{Th} $ and HOMO.
At the scalar ZORA level, electronic state analysis was performed using the natural bond orbital (NBO).
The assumed systems are shown in Fig.~\ref{ADF-model}.
These structures were extracted from the optimized structure to include all the types of the elements
inside the first coordination sphere of $ \mathrm{Th} $ performed in Sec.~\ref{2}.
In all the cases, the charge of $ \mathrm{Th} $ is considered as $ \mathrm{0} $, $ +1 $, $ +2 $, $ +3 $, and $ +4 $.
\begin{table*}[tb]
  \centering
  \caption{Stabilized charge, the number of valence electrons, and the natural orbital configuration of the $ \mathrm{Th} $ atom in the systems calculated.
    We calculated by varying the initial charge of $ \mathrm{Th} $ from $ 0 $ to $ +4 $ and conducted Bader charge analysis for each case.
    In all the calculations, the Bader charge of $ \mathrm{Th} $ remained constant, and this value is presented as stabilized charge.}
  \label{tab:NBO_analysis}
  \begin{ruledtabular}
    \begin{tabular}{lddl}
      \multicolumn{1}{l}{System} & \multicolumn{1}{c}{Stabilized charge} & \multicolumn{1}{c}{The number of valence electrons} & \multicolumn{1}{c}{Natural orbital configuration} \\
      \hline
      $ \mathrm{Ca} \mathrm{F}_2 $ gap
                                 & +2 & 5.03 & $ 7s^{0.93} \, 5f^{0.80} \, 6d^{3.30} $ \\
      $ \mathrm{Ca} \mathrm{F}_2 $-$ \mathrm{Ca} $ replaced
                                 & +3 & 1.63 & $ 7s^{0.19} \, 5f^{0.61} \, 6d^{0.80} $ \\
      $ \mathrm{Mg} \mathrm{F}_2 $ gap
                                 & +2 & 5.54 & $ 7s^{0.63} \, 5f^{0.91} \, 6d^{3.99} $ \\
      $ \mathrm{Mg} \mathrm{F}_2 $-$ \mathrm{Mg} $ replaced
                                 & +3 & 3.24 & $ 7s^{1.88} \, 5f^{0.43} \, 6d^{0.89} $ \\
      $ \mathrm{Li} \mathrm{Sr} \mathrm{Al} \mathrm{F}_6 $ gap
                                 & +1 & 3.73 & $ 7s^{0.75} \, 5f^{0.42} \, 6d^{2.53} $ \\
      $ \mathrm{Li} \mathrm{Sr} \mathrm{Al} \mathrm{F}_6 $-$ \mathrm{Al} $ replaced
                                 & +3 & 2.68 & $ 7s^{0.75} \, 5f^{0.65} \, 6d^{1.09} $ \\
      $ \mathrm{Li} \mathrm{Sr} \mathrm{Al} \mathrm{F}_6 $-$ \mathrm{Sr} $ replaced
                                 & +2 & 3.05 & $ 7s^{0.61} \, 5f^{0.62} \, 6d^{1.81} $ \\
    \end{tabular}
  \end{ruledtabular}
\end{table*}
\begin{table}[tb]
  \centering
  \caption{Binding energy of HOMO of the $ \mathrm{Th} $ atom in the systems calculated.}
  \label{tab:binding_energy}
  \begin{ruledtabular}
    \begin{tabular}{ld}
      \multicolumn{1}{l}{Model} & \multicolumn{1}{c}{The binding energy of the HOMO ($ \mathrm{eV} $)}  \\
      \hline
      $ \mathrm{Ca} \mathrm{F}_2 $ gap
                                & 4.0 \\
      $ \mathrm{Ca} \mathrm{F}_2 $-$ \mathrm{Ca} $ replaced
                                & 3.0 \\
      $ \mathrm{Mg} \mathrm{F}_2 $ gap
                                & 5.9 \\
      $ \mathrm{Mg} \mathrm{F}_2 $-$ \mathrm{Mg} $ replaced
                                & 24.7 \\
      $ \mathrm{Li} \mathrm{Sr} \mathrm{Al} \mathrm{F}_6 $ gap
                                & 15.5 \\
      $ \mathrm{Li} \mathrm{Sr} \mathrm{Al} \mathrm{F}_6 $-$ \mathrm{Al} $ replaced
                                & 4.1 \\
      $ \mathrm{Li} \mathrm{Sr} \mathrm{Al} \mathrm{F}_6 $-$ \mathrm{Sr} $ replaced
                                & 4.3 \\
    \end{tabular}
  \end{ruledtabular}
\end{table}
\subsection{Result and discussion}
\par
The stabilized charge, the number of valence electrons, and the NBOs constituting the valence electrons of the $ \mathrm{Th} $ atom
obtained from band analysis are summarized in Table~\ref{tab:NBO_analysis}. 
The stabilized charge is determined by the Bader charge~\cite{
  RODRIGUEZ2009149} 
and the number of the valence electrons are calculated by sum of the NBO.
It is found that the stabilized charge and the number of valence electrons of $ \mathrm{Th} $ vary depending on the system.
Thus, we performed the following analysis only for cases whose initial charge was identical to the stabilized one.
All the calculations for the initial charge of $ \mathrm{Th} $ show the same stabilized charge.
Here, the initial charge corresponds to the initial state of the charge of $ \mathrm{Th} $ ion before embedding to the sample, while the stabilized charge does the final state after embedding.
The difference of initial and stabilized charge originates from the interactions between the $ \mathrm{Th} $ ion and the other ions surrounding.
\par
The gap model tends to have a larger number of valence electrons than the replaced model in each crystal.
This trend may originate from the fact that the $ \mathrm{Th}$-$ \mathrm{F} $ distance in the gap model is generally shorter than that in the replaced model; 
consequently, the $ \mathrm{Th} $-$ \mathrm{F} $ interaction in the gap model is stronger,
leading to more valence electrons than in the replaced model.
Additionally, the atomic orbitals contributing to the valence electrons in gap models differ from replaced models. 
In the gap model, the contribution from the $ 6d $ orbital is significant,
whereas no such a trend was observed in the replaced model. 
However, this result does not affect the following discussion, since this study focuses on HOMO and $ s $ ones.
In particular, the energy of the HOMO is of the primary importance
no matter what orbitals the HOMO is composed of.
\par
\begin{figure*}
  \centering
  \includegraphics[width=0.8\linewidth]{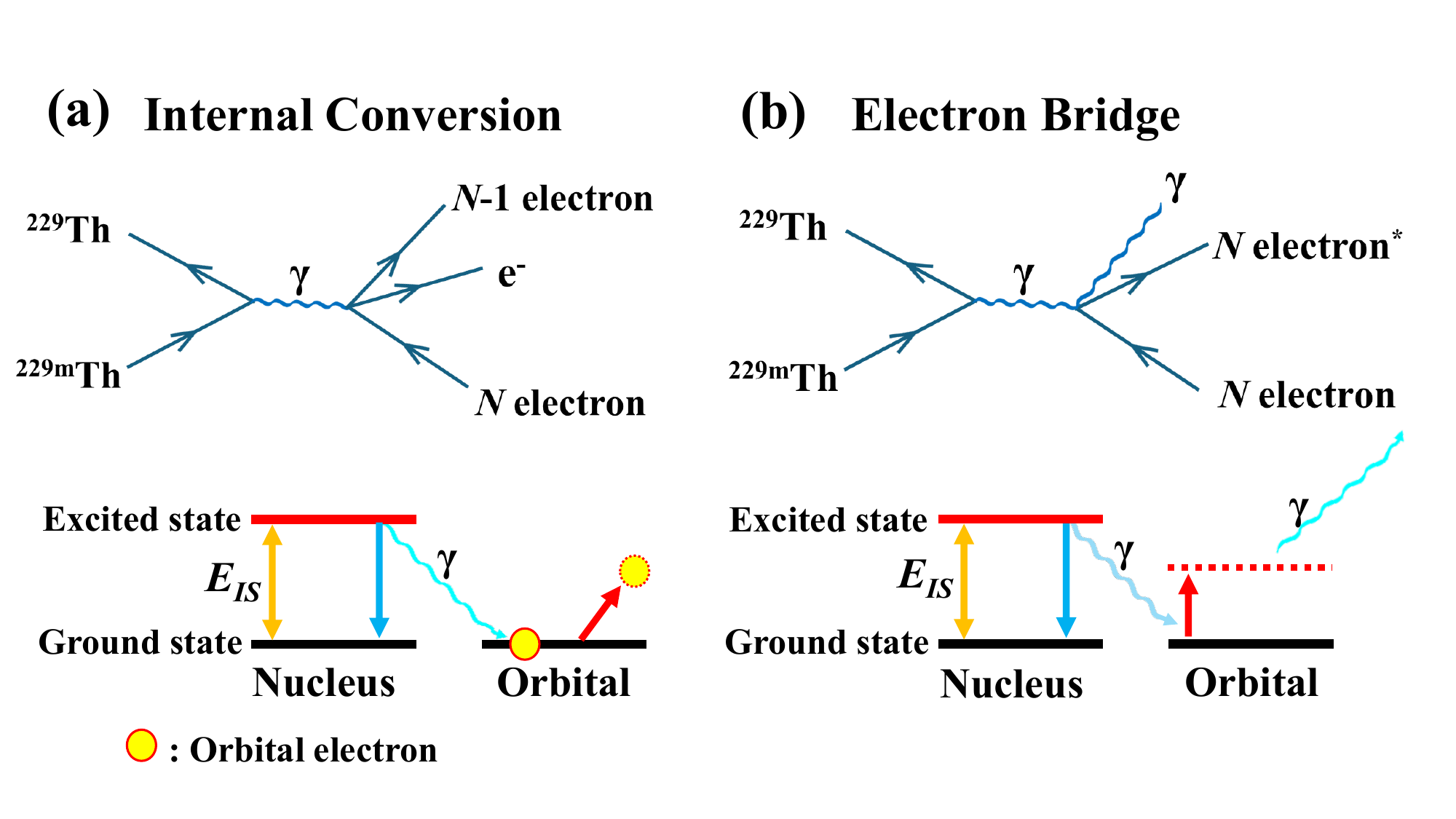}
  \caption{Feynman and schematic diagram of (a) the internal conversion and (b) electron bridge.}
  \label{fig:Feynman_diagram}
\end{figure*}
\begin{figure}
  \centering
  \includegraphics[width=1.0\linewidth]{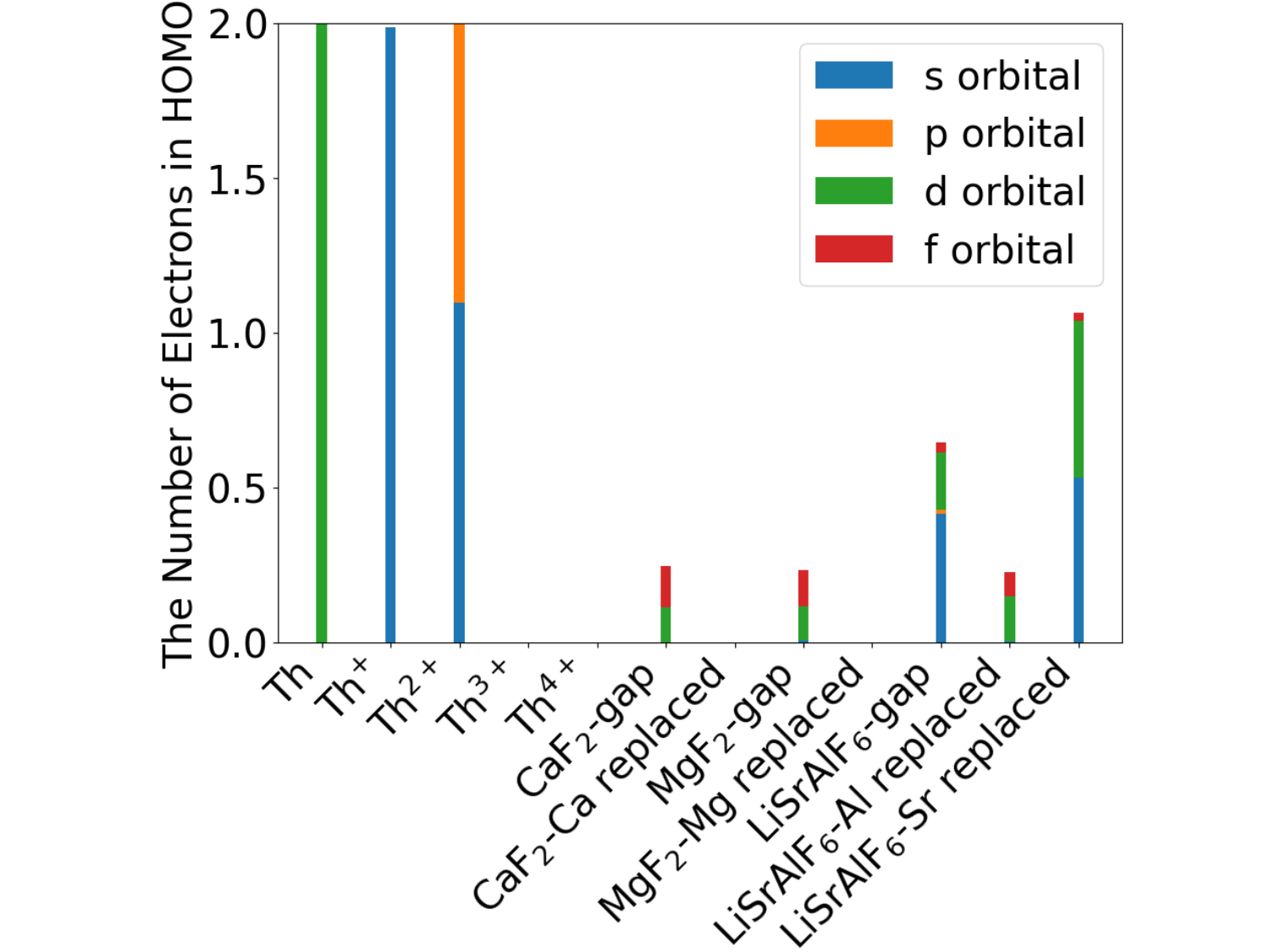}
  \caption{The number of electrons of $ \mathrm{Th} $ that can be involved in IC process and their constituent orbital for each sample.}
  \label{fig:HOMO}
\end{figure}

Next, we focus on the binding energy of the HOMO and $ s $ electron of $ \mathrm{Th} $,
as IC is a process that ejects the outermost electron,
and EB is a process that involved the excitation of $ {s} $ electron into an unoccupied  $ s $ orbital~\cite{
  Karpeshin2018Nucl.Phys.A969_173-183}
(see Fig.~\ref{fig:Feynman_diagram}).
Table~\ref{tab:binding_energy} and Fig.~\ref{fig:HOMO}, respectively, 
show the binding energies of HOMO and the number of electrons that can be emitted as an IC electron and their constituent orbitals in each system.
By comparing the binding energy of HOMO with $ E_{\urm{IS}} $ of $ \nuc{Th}{229m}{} $
($ 8.4 \, \mathrm{eV} $),
we determine IC can occur or not.
In this study, the discussion is carried out under the assumption that only electrons bound to the $ \mathrm{Th} $ atom are allowed to be emitted.
\par
\begin{figure}
  \centering
  \includegraphics[width=1.0\linewidth]{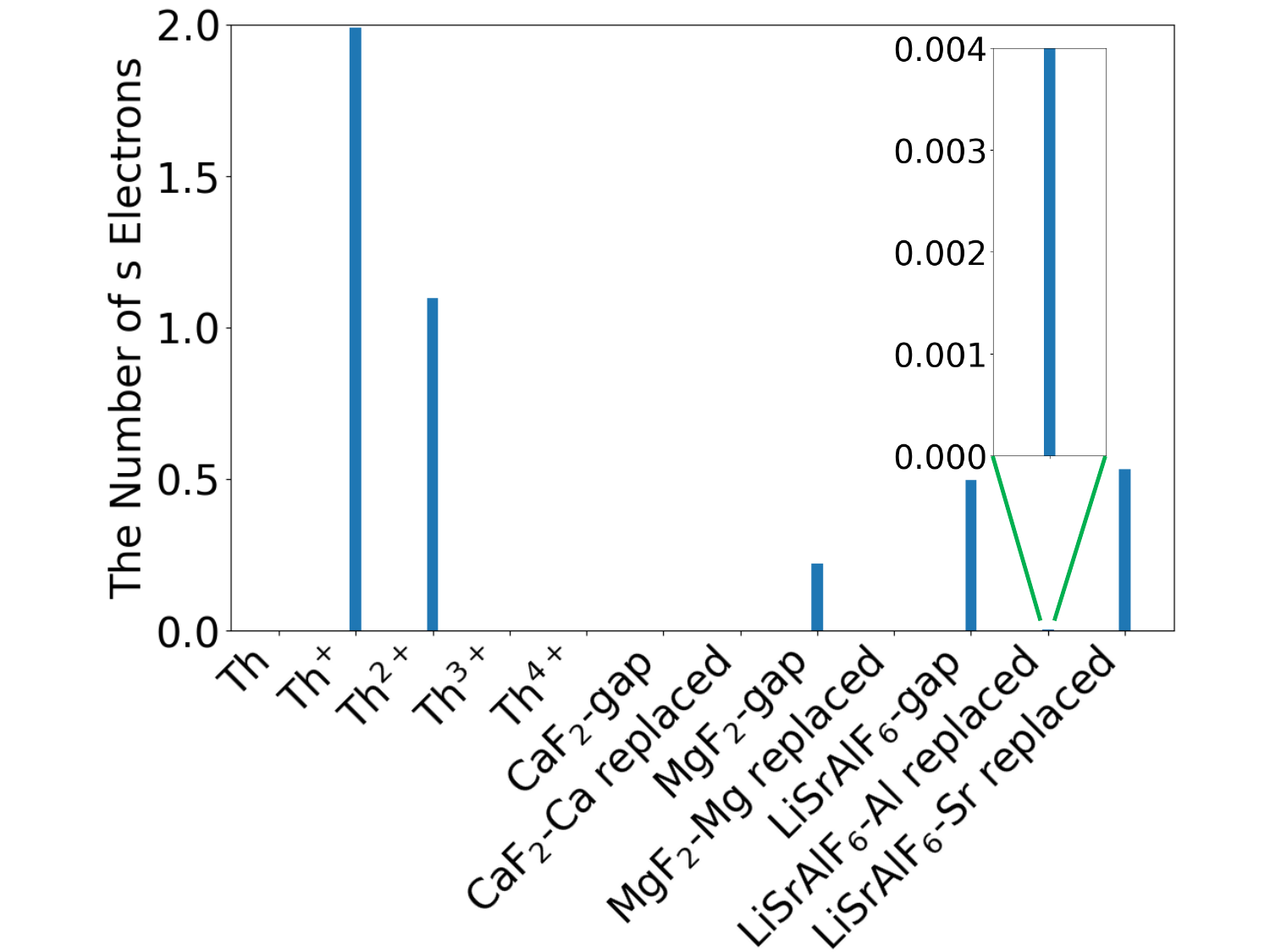}
  \caption{The total number of $ s $ electrons in the levels that are potential initial states for EB.}
  \label{fig:s-electron}
\end{figure}
\begin{figure}
  \centering
  \includegraphics[width=1.0\linewidth]{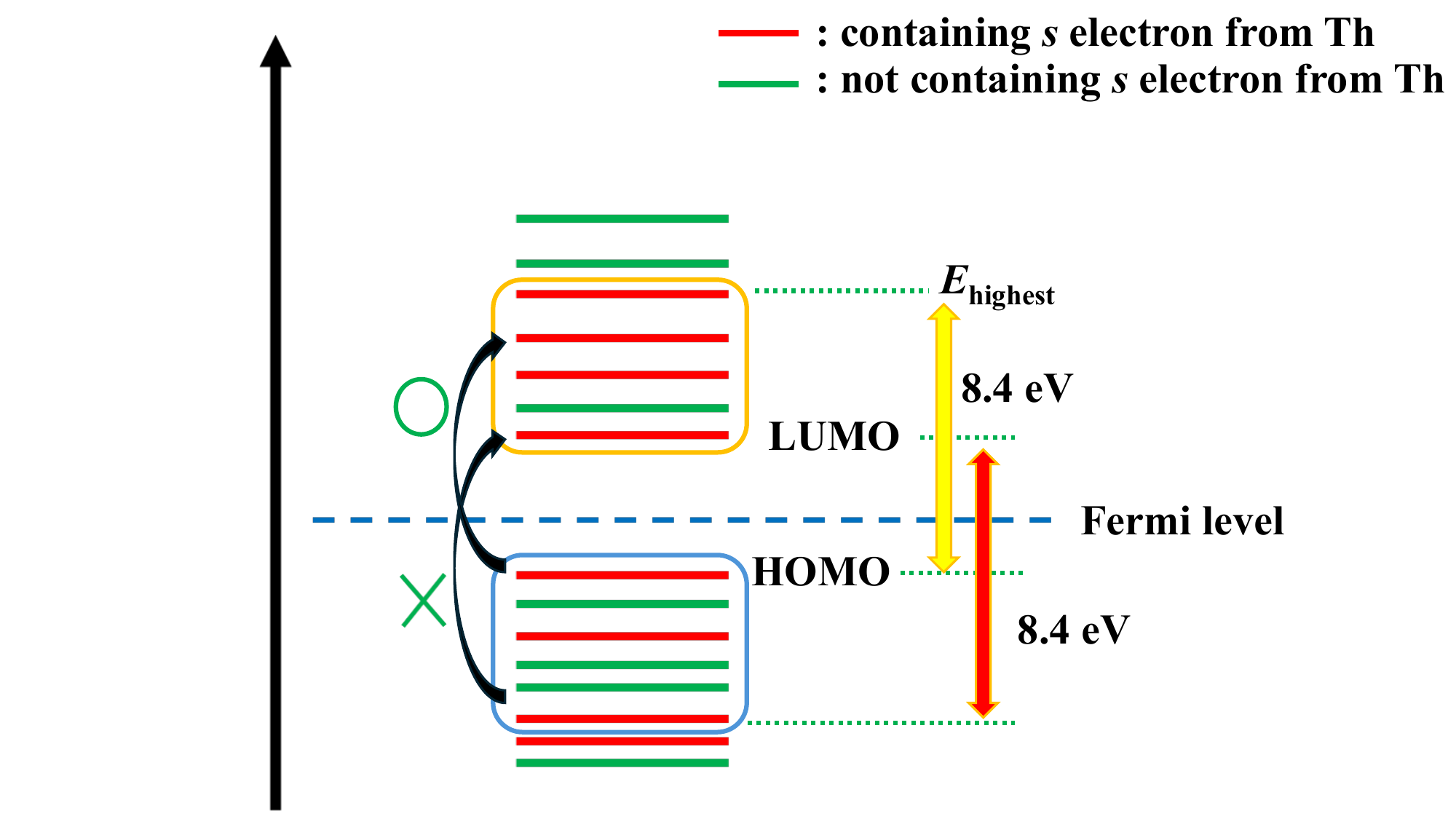}
  \caption{Method for determining the number of $ s $ electrons in the levels involved in EB.}
  \label{fig:how_to_s-electron}
\end{figure}

In case of EB process, as the overlap between $ s $ orbital and the nucleus is large,
this decay mode predominantly involves $ s $ electron from $ \mathrm{Th} $.
Figure~\ref{fig:s-electron} shows the total number of $ s $ electrons of $ \mathrm{Th} $ that involved in EB, which indicates that if $ s $ electrons are present, 
EB is possible. 
This is calculated as follows, which is schematically shown in Fig.~\ref{fig:how_to_s-electron}:
\begin{enumerate}
\item The highest-energy initial state that can contribute to EB is HOMO (the yellow arrow in Fig.~\ref{fig:how_to_s-electron}).
\item The final state should
  include the $ s $ orbital
  and 
  be between the lowest unoccupied molecular orbital and the orbital with $ E_{\urm{highest}} = E_{\urm{HOMO}} + 8.4 \, \mathrm{eV} $,
  where $ E_{\urm{HOMO}} $ denote the energy of HOMO
  (the orange frame in Fig.~\ref{fig:how_to_s-electron}).
\item We determine the lowest-energy initial state
  by subtracting $ 8.4 \, \mathrm{eV} $ from the lowest-energy final state (the red arrow in Fig.~\ref{fig:how_to_s-electron}).
\item We count the number of $ s $ electrons of $ \mathrm{Th} $
  between the lowest initial state and the HOMO (the blue frame in Fig.~\ref{fig:how_to_s-electron}).
\end{enumerate}
\par
In this analysis, the possibility of EB is discussed based on the number of $ s $ electrons in the initial state.
Therefore, the transition probability is not considered, and a larger number of $ s $ electrons does not necessarily mean a higher probability of EB.
\par
The decay constant $ \lambda $ of $ \nuc{Th}{229m}{} $ in crystals was inferred from the reported vacuum half-lives as
\begin{equation}
  \lambda_{\mathrm{Li} \mathrm{Sr} \mathrm{Al} \mathrm{F}_6}
  >
  \lambda_{\mathrm{Ca} \mathrm{F}_2}
  >
  \lambda_{\mathrm{Mg} \mathrm{F}_2}.
\end{equation}
In general, the decay constant is expressed as the sum of the partial decay constants of all the possible decay modes.
We evaluate decay mode of each case by calculation results.
We summarized the calculation results about decay modes in Table~\ref{tab:gamma_ic_eb_analysis}.

\begin{table}[tb]
  \centering
  \caption{Possibility of $ \gamma $-ray transition, IC and EB for each system.
    The circle ($ {\bigcirc} $) and cross ($ {\times} $) symbols correspond to possible and impossible decay mode, respectively. }
  \label{tab:gamma_ic_eb_analysis}
  \begin{ruledtabular}
    \begin{tabular}{llccc}
      Condition & \multicolumn{1}{c}{Previous study} & \multicolumn{3}{c}{Our work} \\
                &                                    & $ \gamma $ & IC & EB \\
      \hline
      Ion trap
                & $ \gamma $-ray transition & $ {\bigcirc} $ & $ {\times} $ & $ {\times} $ \\
      $ \mathrm{Ca} \mathrm{F}_2 $ gap
                & $ \gamma $-ray transition & $ {\bigcirc} $ & $ {\bigcirc} $ & $ {\times} $ \\
      $ \mathrm{Ca} \mathrm{F}_2 $-$\mathrm{Ca}$ replaced
                & $ \gamma $-ray transition & $ {\bigcirc} $ & $ {\times} $ & $ {\times} $ \\
      $ \mathrm{Mg} \mathrm{F}_2 $ gap
                & $ \gamma $-ray transition & $ {\bigcirc} $ & $ {\bigcirc} $ & $ {\bigcirc} $ \\
      $ \mathrm{Mg} \mathrm{F}_2 $-$\mathrm{Mg}$ replaced
                & $ \gamma $-ray transition & $ {\bigcirc} $ & $ {\times} $ & $ {\times} $ \\
      $ \mathrm{Li} \mathrm{Sr}\mathrm{Al}\mathrm{F}_6 $ gap
                & $ \gamma $-ray transition & $ {\bigcirc} $ & $ {\times} $ & $ {\bigcirc} $ \\
      $ \mathrm{Li} \mathrm{Sr}\mathrm{Al}\mathrm{F}_6 $-$\mathrm{Al}$ replaced
                & $ \gamma $-ray transition & $ {\bigcirc} $ & $ {\bigcirc} $ & $ {\bigcirc} $ \\
      $ \mathrm{Li} \mathrm{Sr}\mathrm{Al}\mathrm{F}_6 $-$\mathrm{Sr}$ replaced
                & $ \gamma $-ray transition & $ {\bigcirc} $ & $ {\bigcirc} $ & $ {\bigcirc} $ \\
    \end{tabular}
  \end{ruledtabular}
\end{table}
\subsubsection{$ \nuc{Th}{229m}{} $ atom and ions}
\par
In case of the neutral $ \nuc{Th}{229m}{} $, the binding energy of HOMO is lower than $ E_{\urm{IS}} $;
thus, IC can occur.
There is no electron involved in EB; thus, EB is forbidden.
On the other hand, in case of $ \nuc{Th}{229m}{}^{{+}, \, {2+}, \, {3+}, \, {4+}} $, the binding energy of HOMO is lager than the $ E_{\urm{IS}} = 8.4 \, \mathrm{eV} $, and therefore IC is forbidden.
In addition,
according to Fig.~\ref{fig:s-electron},
$ \nuc{Th}{229m}{}^{{3+}, \, {4+}} $ do not have $ s $ electrons,
which can be involved in EB,
while $ \nuc{Th}{229m}{}^{{+}, \, {2+}} $ do.
Therefore, the decay constant is expressed as
\begin{subequations}
  \begin{align}
    \lambda_{\mathrm{Th}}
    & =
      \lambda_{\gamma}
      +
      \lambda_{\urm{IC}}, \\
    \lambda_{\mathrm{Th}^{{+}, \, {2+}}}
    & =
      \lambda_{\gamma}
      +
      \lambda_{\urm{EB}}, \\
    \lambda_{\mathrm{Th}^{{3+}, \, {4+}}}
    & =
      \lambda_{\gamma},
  \end{align}
\end{subequations}
where we assume that the neutral $ \nuc{Th}{229m}{} $ atom can decay via the $ \gamma $ decay as well,
since the $ \gamma $ decay is, in principle, independent from the environment.
\par
In the reported experiment based on the ion trap, $ \nuc{Th}{229m}{}^{3+} $ was kept.
It is known that IC has shorter half-life than $ \gamma $-ray transition~\cite{PhysRevLett.118.042501},
and a calculation predicts that EB has shorter half-life than the $ \gamma $-ray transition~\cite{Karpeshin2018Nucl.Phys.A969_173-183}.
Thus, the decay constant of the ion trap method is expected to be the smallest among the decay constants in vacuum, 
which agrees with experimental results within the error.
This indicates that ion trap method, $ \nuc{Th}{229m}{}^{3+} $, decays only via $ \gamma $-ray transition.
\par
Focusing the calculation results of $ \nuc{Th}{229m}{}^{{+}, \, {2+}, \, {4+}} $, it is indicated that the half-lives of $ \nuc{Th}{229m}{}^{{+}, \, {2+}} $ may be shorter than that of $ \nuc{Th}{229m}{}^{3+} $.
It was reported that the half-lives of $ \nuc{Th}{229m}{}^{{+}, \, {2+}} $ is longer than $ 60 \, \mathrm{s} $~\cite{
  VonDerWenseNature553_042501}, 
and this reported value is consistent with calculation results.
The half-life of $ \nuc{Th}{229m}{}^{4+} $ is comparable to $ \nuc{Th}{229m}{}^{3+} $.
In case of the neutral $ \nuc{Th}{229m}{} $, 
the half-life is experimentally reported to be $ 7 \, \mathrm{\mu s} $, and the decay constant $ \lambda_{\urm{IC}} $ is nine orders of magnitude larger than $ \lambda_{\gamma} $~\cite{
  Yamaguchi2024Nature.629_62,
  VonDerWenseNature553_042501}.
Thus, it is considered that IC is the dominating decay mode.
A detailed discussion requires the probabilities of IC through comparison with experimental data, which is left for future work.
\subsubsection{$ \mathrm{Ca} \mathrm{F}_2 $}
\par
The binding energy of HOMO is lower than $ E_{\urm{IS}} = 8.4 \, \mathrm{eV} $
in both gap and replaced models,
indicating that IC can occur.
Focusing on the replaced model,
electrons that can de emitted do not exist in the $ \mathrm{Th} $ orbital (see Fig.~\ref{fig:HOMO}).
Hence, IC is energetically possible based on the binding energy of HOMO, but there is no electron involved in IC.
Therefore, this sample is considered as IC forbidden.
There is no electron involved in EB in both models (see Fig.~\ref{fig:s-electron});
thus, it cannot decay via EB.
Thus, the decay constants for each case are expressed as
\begin{subequations}
  \begin{align}
    \lambda_{\urm{$ \mathrm{Ca} \mathrm{F}_2 $ gap}}
    & = 
      \lambda_{\gamma}
      +
      \lambda_{\urm{IC}}, \\
    \lambda_{\urm{$ \mathrm{Ca} \mathrm{F}_2 $ replaced}}
    & =
      \lambda_{\gamma}.
  \end{align}
\end{subequations}
\par
Previous study has reported that the experiment was conducted in replaced model~\cite{
  Dessovic_2014}.
This suggests that $ \nuc{Th}{229m}{} $ in $ \mathrm{Ca} \mathrm{F}_2 $ decays only via the $ \gamma $-ray transition.
Our analysis supports the experimental result,
as the reported half-life is consistent with that obtained by the ion trap method within the error.
\subsubsection{$ \mathrm{Mg} \mathrm{F}_2 $}
\par
In the gap model, the binding energy of HOMO is lower than $ E_{\urm{IS}} = 8.4 \, \mathrm{eV} $, leading to IC being possible, and from Fig.~\ref{fig:s-electron}, 
there are electrons involved in EB.
On the other hand, in the replaced model, the binding energy of HOMO is larger than $ E_{\urm{IS}} $, leading to IC being forbidden,
and from Fig.~\ref{fig:s-electron}, there is no electron involved in EB. 
Thus, the decay constants for each model are expressed as
\begin{subequations}
  \begin{align}
    \lambda_{\urm{$ \mathrm{Mg} \mathrm{F}_2 $ gap}}
    & = 
      \lambda_{\gamma}
      +
      \lambda_{\urm{IC}}
      +
      \lambda_{\urm{EB}}, \\
    \lambda_{\urm{$ \mathrm{Mg} \mathrm{F}_2 $ replaced}}
    & =
      \lambda_{\gamma}.
  \end{align}
\end{subequations}
\par
The structure of experimental sample, whether gap or replaced model, remains unclear.
Considering the reported half-lives relationship
$ T_{\mathrm{Ca} \mathrm{F}_2} < T_{\mathrm{Mg} \mathrm{F}_2} $,
$ \mathrm{Mg}\mathrm{F}_2 $ model does not decay via IC and EB as such decay modes would result in a significantly shorter half-life.
Hence, we conclude that the replaced model is appropriate and the decay constant is
$ \lambda_{\mathrm{Mg} \mathrm{F}_2} = \lambda_{\gamma} $.
However, there is a difference in half-lives between $ \mathrm{Ca}\mathrm{F}_2 $ and $ \mathrm{Mg}\mathrm{F}_2 $, which may originate the incorrect refractive index multiplied in the conversion process from the half-life in crystal to that in vacuum.
\subsubsection{$ \mathrm{Li} \mathrm{Sr} \mathrm{Al} \mathrm{F}_6 $}
\par
In the gap model, the binding energy of HOMO is larger than $ E_{\urm{IS}} $, leading to IC being forbidden, and from Fig.~\ref{fig:s-electron}, there are electrons involved EB.
On the other hand, in the replaced model, the binding energy of HOMO is lower than $ E_{\urm{IS}} $,
leading to IC being possible, and from Fig.~\ref{fig:s-electron}, there are electrons involved in EB. 
Thus, the decay constant for each model is expressed as
\begin{subequations}
  \begin{align}
    \lambda_{\urm{$ \mathrm{Li} \mathrm{Sr} \mathrm{Al} \mathrm{F}_6 $ gap}}
    & =
      \lambda_{\gamma}
      +
      \lambda_{\urm{EB}}, \\
    \lambda_{\urm{$ \mathrm{Li} \mathrm{Sr} \mathrm{Al} \mathrm{F}_6 $ replaced}}
    & =
      \lambda_{\gamma}
      +
      \lambda_{\urm{IC}}
      +
      \lambda_{\urm{EB}}.
  \end{align}
\end{subequations}
\par
Similar to the $ \mathrm{Mg} \mathrm{F}_2 $ case, it is unclear whether the chemical structure is the gap or replaced model.
In both models, the $ \gamma $-ray transition is not the unique decay mode. 
Therefore, the half-life should be shorter than ion trap. 
These results are consistent with the reported relationship of half-lives 
($ T_{\mathrm{Li} \mathrm{Sr} \mathrm{Al} \mathrm{F}_6} < T_{\mathrm{Ca} \mathrm{F}_2} $).
\section{Summary}
\par
We performed electronic structure analysis of $ \mathrm{Th} $ using 
\textit{ab initio} calculations
to understand the discrepancies among the reported half-lives of $ \nuc{Th}{229m}{} $ of the $ \gamma $-ray transition.
We successfully conducted, for the first time, the qualitative evaluation of the decay mode of $ \nuc{Th}{229m}{} $.
This achievement is expected to contribute determining the most suitable candidate of implementing a nuclear clock and estimate the placement of $ \mathrm{Th} $ in each crystal.
\par
The decay modes are different depending on the systems due to the differences of electronic state.
In cases of the ion trap method, $ \mathrm{Ca} \mathrm{F}_2 $, and $ \mathrm{Mg} \mathrm{F}_2 $,
the decay mode is only the $ \gamma $-ray transition.
On the other hand, in case of $ \mathrm{Li} \mathrm{Sr} \mathrm{Al} \mathrm{F}_6 $,
the decay modes are not only the $ \gamma $-ray transition but also internal conversion and electron bridge.
These results suggest that the half-life of $ \mathrm{Li} \mathrm{Sr} \mathrm{Al} \mathrm{F}_6 $ is shorter than the half-lives of the other systems discussed,
but do not explain the discrepancies among the half-lives of the other cases.
The reaming discrepancies may originate from the changing refractive index around the $ \mathrm{Th} $ ion,
which is left for a future perspective.
The probability of the internal conversion should also be estimated toward quantitative discussion;
we plan to analyze it quantitatively with a microscopic calculation of the nuclear structure.
\begin{acknowledgments}
  This work was partly achieved through the use of SQUID at D3 Center, Osaka University.
  This work was supported by JSPS KAKENHI Grant Nos.~JP22K20372,
  JP23H01845,
  JP23H04526,
  JP23K03426,
  JP23K04636,
  JP23K26538,
  JP24K17057,
  JP25H00402,
  JP25H01558,
  JP25K01003, and
  JP25KJ0405
  and
  JST COI-NEXT Grant No.~JPMJPF2221.
  T.~N.~acknolwedges the RIKEN Special Postdoctoral Researcher Program. 
  We acknowledge the VESTA software~\cite{
    Momma:db5098}
  for the visualization in Figs.~\ref{QE-model} and \ref{ADF-model}.
\end{acknowledgments}

\begin{thebibliography}{41}%
\makeatletter
\providecommand \@ifxundefined [1]{%
 \@ifx{#1\undefined}
}%
\providecommand \@ifnum [1]{%
 \ifnum #1\expandafter \@firstoftwo
 \else \expandafter \@secondoftwo
 \fi
}%
\providecommand \@ifx [1]{%
 \ifx #1\expandafter \@firstoftwo
 \else \expandafter \@secondoftwo
 \fi
}%
\providecommand \natexlab [1]{#1}%
\providecommand \enquote  [1]{``#1''}%
\providecommand \bibnamefont  [1]{#1}%
\providecommand \bibfnamefont [1]{#1}%
\providecommand \citenamefont [1]{#1}%
\providecommand \href@noop [0]{\@secondoftwo}%
\providecommand \href [0]{\begingroup \@sanitize@url \@href}%
\providecommand \@href[1]{\@@startlink{#1}\@@href}%
\providecommand \@@href[1]{\endgroup#1\@@endlink}%
\providecommand \@sanitize@url [0]{\catcode `\\12\catcode `\$12\catcode
  `\&12\catcode `\#12\catcode `\^12\catcode `\_12\catcode `\%12\relax}%
\providecommand \@@startlink[1]{}%
\providecommand \@@endlink[0]{}%
\providecommand \url  [0]{\begingroup\@sanitize@url \@url }%
\providecommand \@url [1]{\endgroup\@href {#1}{\urlprefix }}%
\providecommand \urlprefix  [0]{URL }%
\providecommand \Eprint [0]{\href }%
\providecommand \doibase [0]{https://doi.org/}%
\providecommand \selectlanguage [0]{\@gobble}%
\providecommand \bibinfo  [0]{\@secondoftwo}%
\providecommand \bibfield  [0]{\@secondoftwo}%
\providecommand \translation [1]{[#1]}%
\providecommand \BibitemOpen [0]{}%
\providecommand \bibitemStop [0]{}%
\providecommand \bibitemNoStop [0]{.\EOS\space}%
\providecommand \EOS [0]{\spacefactor3000\relax}%
\providecommand \BibitemShut  [1]{\csname bibitem#1\endcsname}%
\let\auto@bib@innerbib\@empty
%</preamble>
\bibitem [{\citenamefont {Ponce}\ \emph {et~al.}(2018)\citenamefont {Ponce},
  \citenamefont {Swanberg}, \citenamefont {Burke}, \citenamefont {Henderson},\
  and\ \citenamefont {Friedrich}}]{PhysRevC.97.054310}%
  \BibitemOpen
  \bibfield  {author} {\bibinfo {author} {\bibfnamefont {F.}~\bibnamefont
  {Ponce}}, \bibinfo {author} {\bibfnamefont {E.}~\bibnamefont {Swanberg}},
  \bibinfo {author} {\bibfnamefont {J.}~\bibnamefont {Burke}}, \bibinfo
  {author} {\bibfnamefont {R.}~\bibnamefont {Henderson}},\ and\ \bibinfo
  {author} {\bibfnamefont {S.}~\bibnamefont {Friedrich}},\ }\bibfield  {title}
  {\bibinfo {title} {{Accurate measurement of the first excited nuclear state
  in $^{235}\mathrm{U}$}},\ }\href {https://doi.org/10.1103/PhysRevC.97.054310}
  {\bibfield  {journal} {\bibinfo  {journal} {Phys. Rev. C}\ }\textbf {\bibinfo
  {volume} {97}},\ \bibinfo {pages} {054310} (\bibinfo {year}
  {2018})}\BibitemShut {NoStop}%
\bibitem [{\citenamefont
  {de~Mevergnies}(1972)}]{Mevergnies1972Phys.Rev.Lett.29_1188}%
  \BibitemOpen
  \bibfield  {author} {\bibinfo {author} {\bibfnamefont {M.~N.}\ \bibnamefont
  {de~Mevergnies}},\ }\bibfield  {title} {\bibinfo {title} {{Perturbation of
  the $^{235m}\mathrm{U}$ Decay Rate by Implantation in Transition Metals}},\
  }\href {https://doi.org/10.1103/PhysRevLett.29.1188} {\bibfield  {journal}
  {\bibinfo  {journal} {Phys. Rev. Lett.}\ }\textbf {\bibinfo {volume} {29}},\
  \bibinfo {pages} {1188} (\bibinfo {year} {1972})}\BibitemShut {NoStop}%
\bibitem [{\citenamefont {de~Mevergnies}\ and\ \citenamefont
  {Del~Marmol}(1974)}]{Mevergnies1974Phys.Lett.B49_428}%
  \BibitemOpen
  \bibfield  {author} {\bibinfo {author} {\bibfnamefont {M.~N.}\ \bibnamefont
  {de~Mevergnies}}\ and\ \bibinfo {author} {\bibfnamefont {P.}~\bibnamefont
  {Del~Marmol}},\ }\bibfield  {title} {\bibinfo {title} {{Effect of the
  oxidation state on the half-life of $ {}^{235} \mathrm{U}^{\mathrm{m}} $}},\
  }\href {https://doi.org/10.1016/0370-2693(74)90626-1} {\bibfield  {journal}
  {\bibinfo  {journal} {Phys. Lett. B}\ }\textbf {\bibinfo {volume} {49}},\
  \bibinfo {pages} {428} (\bibinfo {year} {1974})}\BibitemShut {NoStop}%
\bibitem [{\citenamefont {Kraemer}(2023)}]{Kremar2023Nature.617_706}%
  \BibitemOpen
  \bibfield  {author} {\bibinfo {author} {\bibfnamefont {S.}~\bibnamefont
  {Kraemer}},\ }\bibfield  {title} {\bibinfo {title} {{Observation of the
  radiative decay of the $ {}^{229} \mathrm{Th}^{\mathrm{m}} $ nuclear clock
  isomer}},\ }\href {https://doi.org/10.1038/s41586-023-05894-z} {\bibfield
  {journal} {\bibinfo  {journal} {Nature}\ }\textbf {\bibinfo {volume} {617}},\
  \bibinfo {pages} {706} (\bibinfo {year} {2023})}\BibitemShut {NoStop}%
\bibitem [{\citenamefont {Yamaguchi}\ \emph {et~al.}(2024)\citenamefont
  {Yamaguchi}, \citenamefont {Shigekawa}, \citenamefont {Haba}, \citenamefont
  {Kikunaga}, \citenamefont {Shirasaki}, \citenamefont {Wada},\ and\
  \citenamefont {Katori}}]{Yamaguchi2024Nature.629_62}%
  \BibitemOpen
  \bibfield  {author} {\bibinfo {author} {\bibfnamefont {A.}~\bibnamefont
  {Yamaguchi}}, \bibinfo {author} {\bibfnamefont {Y.}~\bibnamefont
  {Shigekawa}}, \bibinfo {author} {\bibfnamefont {H.}~\bibnamefont {Haba}},
  \bibinfo {author} {\bibfnamefont {H.}~\bibnamefont {Kikunaga}}, \bibinfo
  {author} {\bibfnamefont {K.}~\bibnamefont {Shirasaki}}, \bibinfo {author}
  {\bibfnamefont {M.}~\bibnamefont {Wada}},\ and\ \bibinfo {author}
  {\bibfnamefont {H.}~\bibnamefont {Katori}},\ }\bibfield  {title} {\bibinfo
  {title} {{Laser spectroscopy of triply charged $ {}^{229} \mathrm{Th} $
  isomer for a nuclear clock}},\ }\href
  {https://doi.org/10.1038/s41586-024-07296-1} {\bibfield  {journal} {\bibinfo
  {journal} {Nature}\ }\textbf {\bibinfo {volume} {629}},\ \bibinfo {pages}
  {62} (\bibinfo {year} {2024})}\BibitemShut {NoStop}%
\bibitem [{\citenamefont {Tiedau}\ \emph {et~al.}(2024)\citenamefont {Tiedau},
  \citenamefont {Okhapkin}, \citenamefont {Zhang}, \citenamefont {Thielking},
  \citenamefont {Zitzer}, \citenamefont {Peik}, \citenamefont {Schaden},
  \citenamefont {Pronebner}, \citenamefont {Morawetz}, \citenamefont {De~Col},
  \citenamefont {Schneider}, \citenamefont {Leitner}, \citenamefont {Pressler},
  \citenamefont {Kazakov}, \citenamefont {Beeks}, \citenamefont {Sikorsky},\
  and\ \citenamefont {Schumm}}]{Tiedau2024Phys.Rev.Lett.132_182501}%
  \BibitemOpen
  \bibfield  {author} {\bibinfo {author} {\bibfnamefont {J.}~\bibnamefont
  {Tiedau}}, \bibinfo {author} {\bibfnamefont {M.~V.}\ \bibnamefont
  {Okhapkin}}, \bibinfo {author} {\bibfnamefont {K.}~\bibnamefont {Zhang}},
  \bibinfo {author} {\bibfnamefont {J.}~\bibnamefont {Thielking}}, \bibinfo
  {author} {\bibfnamefont {G.}~\bibnamefont {Zitzer}}, \bibinfo {author}
  {\bibfnamefont {E.}~\bibnamefont {Peik}}, \bibinfo {author} {\bibfnamefont
  {F.}~\bibnamefont {Schaden}}, \bibinfo {author} {\bibfnamefont
  {T.}~\bibnamefont {Pronebner}}, \bibinfo {author} {\bibfnamefont
  {I.}~\bibnamefont {Morawetz}}, \bibinfo {author} {\bibfnamefont {L.~T.}\
  \bibnamefont {De~Col}}, \bibinfo {author} {\bibfnamefont {F.}~\bibnamefont
  {Schneider}}, \bibinfo {author} {\bibfnamefont {A.}~\bibnamefont {Leitner}},
  \bibinfo {author} {\bibfnamefont {M.}~\bibnamefont {Pressler}}, \bibinfo
  {author} {\bibfnamefont {G.~A.}\ \bibnamefont {Kazakov}}, \bibinfo {author}
  {\bibfnamefont {K.}~\bibnamefont {Beeks}}, \bibinfo {author} {\bibfnamefont
  {T.}~\bibnamefont {Sikorsky}},\ and\ \bibinfo {author} {\bibfnamefont
  {T.}~\bibnamefont {Schumm}},\ }\bibfield  {title} {\bibinfo {title} {{Laser
  Excitation of the Th-229 Nucleus}},\ }\href
  {https://doi.org/10.1103/PhysRevLett.132.182501} {\bibfield  {journal}
  {\bibinfo  {journal} {Phys. Rev. Lett.}\ }\textbf {\bibinfo {volume} {132}},\
  \bibinfo {pages} {182501} (\bibinfo {year} {2024})}\BibitemShut {NoStop}%
\bibitem [{\citenamefont {Elwell}\ \emph {et~al.}(2024)\citenamefont {Elwell},
  \citenamefont {Schneider}, \citenamefont {Jeet}, \citenamefont {Terhune},
  \citenamefont {Morgan}, \citenamefont {Alexandrova}, \citenamefont
  {Tran~Tan}, \citenamefont {Derevianko},\ and\ \citenamefont
  {Hudson}}]{Elwell2024Phys.Rev.Lett.133_013201}%
  \BibitemOpen
  \bibfield  {author} {\bibinfo {author} {\bibfnamefont {R.}~\bibnamefont
  {Elwell}}, \bibinfo {author} {\bibfnamefont {C.}~\bibnamefont {Schneider}},
  \bibinfo {author} {\bibfnamefont {J.}~\bibnamefont {Jeet}}, \bibinfo {author}
  {\bibfnamefont {J.~E.~S.}\ \bibnamefont {Terhune}}, \bibinfo {author}
  {\bibfnamefont {H.~W.~T.}\ \bibnamefont {Morgan}}, \bibinfo {author}
  {\bibfnamefont {A.~N.}\ \bibnamefont {Alexandrova}}, \bibinfo {author}
  {\bibfnamefont {H.~B.}\ \bibnamefont {Tran~Tan}}, \bibinfo {author}
  {\bibfnamefont {A.}~\bibnamefont {Derevianko}},\ and\ \bibinfo {author}
  {\bibfnamefont {E.~R.}\ \bibnamefont {Hudson}},\ }\bibfield  {title}
  {\bibinfo {title} {{Laser Excitation of the $^{229}\mathrm{Th}$ Nuclear
  Isomeric Transition in a Solid-State Host}},\ }\href
  {https://doi.org/10.1103/PhysRevLett.133.013201} {\bibfield  {journal}
  {\bibinfo  {journal} {Phys. Rev. Lett.}\ }\textbf {\bibinfo {volume} {133}},\
  \bibinfo {pages} {013201} (\bibinfo {year} {2024})}\BibitemShut {NoStop}%
\bibitem [{\citenamefont {Hiraki}\ \emph {et~al.}(2024)\citenamefont {Hiraki},
  \citenamefont {Okai}, \citenamefont {Bartokos}, \citenamefont {Beeks},
  \citenamefont {Fujimoto}, \citenamefont {Fukunaga}, \citenamefont {Haba},
  \citenamefont {Kasamatsu}, \citenamefont {Kitao}, \citenamefont {Leitner},
  \citenamefont {Masuda}, \citenamefont {Guan}, \citenamefont {Nagasawa},
  \citenamefont {Ogake}, \citenamefont {Pimon}, \citenamefont {Pressler},
  \citenamefont {Sasao}, \citenamefont {Schaden}, \citenamefont {Schumm},
  \citenamefont {Seto}, \citenamefont {Shigekawa}, \citenamefont {Shimizu},
  \citenamefont {Sikorsky}, \citenamefont {Tamasaku}, \citenamefont {Takatori},
  \citenamefont {Watanabe}, \citenamefont {Yamaguchi}, \citenamefont {Yoda},
  \citenamefont {Yoshimi},\ and\ \citenamefont
  {Yoshimura}}]{Hiraki2024Nat.Commun.15_5536}%
  \BibitemOpen
  \bibfield  {author} {\bibinfo {author} {\bibfnamefont {T.}~\bibnamefont
  {Hiraki}}, \bibinfo {author} {\bibfnamefont {K.}~\bibnamefont {Okai}},
  \bibinfo {author} {\bibfnamefont {M.}~\bibnamefont {Bartokos}}, \bibinfo
  {author} {\bibfnamefont {K.}~\bibnamefont {Beeks}}, \bibinfo {author}
  {\bibfnamefont {H.}~\bibnamefont {Fujimoto}}, \bibinfo {author}
  {\bibfnamefont {Y.}~\bibnamefont {Fukunaga}}, \bibinfo {author}
  {\bibfnamefont {H.}~\bibnamefont {Haba}}, \bibinfo {author} {\bibfnamefont
  {Y.}~\bibnamefont {Kasamatsu}}, \bibinfo {author} {\bibfnamefont
  {S.}~\bibnamefont {Kitao}}, \bibinfo {author} {\bibfnamefont
  {A.}~\bibnamefont {Leitner}}, \bibinfo {author} {\bibfnamefont
  {T.}~\bibnamefont {Masuda}}, \bibinfo {author} {\bibfnamefont
  {M.}~\bibnamefont {Guan}}, \bibinfo {author} {\bibfnamefont {N.}~\bibnamefont
  {Nagasawa}}, \bibinfo {author} {\bibfnamefont {R.}~\bibnamefont {Ogake}},
  \bibinfo {author} {\bibfnamefont {M.}~\bibnamefont {Pimon}}, \bibinfo
  {author} {\bibfnamefont {M.}~\bibnamefont {Pressler}}, \bibinfo {author}
  {\bibfnamefont {N.}~\bibnamefont {Sasao}}, \bibinfo {author} {\bibfnamefont
  {F.}~\bibnamefont {Schaden}}, \bibinfo {author} {\bibfnamefont
  {T.}~\bibnamefont {Schumm}}, \bibinfo {author} {\bibfnamefont
  {M.}~\bibnamefont {Seto}}, \bibinfo {author} {\bibfnamefont {Y.}~\bibnamefont
  {Shigekawa}}, \bibinfo {author} {\bibfnamefont {K.}~\bibnamefont {Shimizu}},
  \bibinfo {author} {\bibfnamefont {T.}~\bibnamefont {Sikorsky}}, \bibinfo
  {author} {\bibfnamefont {K.}~\bibnamefont {Tamasaku}}, \bibinfo {author}
  {\bibfnamefont {S.}~\bibnamefont {Takatori}}, \bibinfo {author}
  {\bibfnamefont {T.}~\bibnamefont {Watanabe}}, \bibinfo {author}
  {\bibfnamefont {A.}~\bibnamefont {Yamaguchi}}, \bibinfo {author}
  {\bibfnamefont {Y.}~\bibnamefont {Yoda}}, \bibinfo {author} {\bibfnamefont
  {A.}~\bibnamefont {Yoshimi}},\ and\ \bibinfo {author} {\bibfnamefont
  {K.}~\bibnamefont {Yoshimura}},\ }\bibfield  {title} {\bibinfo {title}
  {{Controlling $ {}^{229} \mathrm{Th} $ isomeric state population in a VUV
  transparent crystal}},\ }\href {https://doi.org/10.1038/s41467-024-49631-0}
  {\bibfield  {journal} {\bibinfo  {journal} {Nat. Commun.}\ }\textbf {\bibinfo
  {volume} {15}},\ \bibinfo {pages} {5536} (\bibinfo {year}
  {2024})}\BibitemShut {NoStop}%
\bibitem [{\citenamefont {Strizhov}\ and\ \citenamefont
  {Tkalya}(1991)}]{Strizhov1991}%
  \BibitemOpen
  \bibfield  {author} {\bibinfo {author} {\bibfnamefont {V.~F.}\ \bibnamefont
  {Strizhov}}\ and\ \bibinfo {author} {\bibfnamefont {E.~V.}\ \bibnamefont
  {Tkalya}},\ }\bibfield  {title} {\bibinfo {title} {{Decay channel of
  low-lying isomer state of the $^{229}$Th nucleus. Possibilities of
  experimental investigation}},\ }\href@noop {} {\bibfield  {journal} {\bibinfo
   {journal} {Zh. Eksp. Teor. Fiz.}\ }\textbf {\bibinfo {volume} {99}},\
  \bibinfo {pages} {697} (\bibinfo {year} {1991})},\ \bibinfo {note} {[Sov.
  Phys. JETP \textbf{72}, 387 (1991)]}\BibitemShut {NoStop}%
\bibitem [{\citenamefont {Dzuba}\ and\ \citenamefont
  {Flambaum}(2025{\natexlab{a}})}]{PhysRevA.111.L041103}%
  \BibitemOpen
  \bibfield  {author} {\bibinfo {author} {\bibfnamefont {V.~A.}\ \bibnamefont
  {Dzuba}}\ and\ \bibinfo {author} {\bibfnamefont {V.~V.}\ \bibnamefont
  {Flambaum}},\ }\bibfield  {title} {\bibinfo {title} {{Resonance nuclear
  excitation of the $^{229}\mathrm{Th}$ nucleus via electronic bridge process
  in Th ii}},\ }\href {https://doi.org/10.1103/PhysRevA.111.L041103} {\bibfield
   {journal} {\bibinfo  {journal} {Phys. Rev. A}\ }\textbf {\bibinfo {volume}
  {111}},\ \bibinfo {pages} {L041103} (\bibinfo {year}
  {2025}{\natexlab{a}})}\BibitemShut {NoStop}%
\bibitem [{\citenamefont {Karpeshin}\ and\ \citenamefont
  {Trzhaskovskaya}(2018)}]{Karpeshin2018Nucl.Phys.A969_173-183}%
  \BibitemOpen
  \bibfield  {author} {\bibinfo {author} {\bibfnamefont {F.~E.}\ \bibnamefont
  {Karpeshin}}\ and\ \bibinfo {author} {\bibfnamefont {M.~B.}\ \bibnamefont
  {Trzhaskovskaya}},\ }\bibfield  {title} {\bibinfo {title} {{Impact of the
  ionization of the atomic shell on the lifetime of the $ {}^{229} \mathrm{Th}
  $ isomer}},\ }\href {https://doi.org/10.1016/j.nuclphysa.2017.10.003}
  {\bibfield  {journal} {\bibinfo  {journal} {Nucl. Phys. A}\ }\textbf
  {\bibinfo {volume} {969}},\ \bibinfo {pages} {173} (\bibinfo {year}
  {2018})}\BibitemShut {NoStop}%
\bibitem [{\citenamefont {von~der Wense}\ and\ \citenamefont
  {Seiferle}(2020)}]{VonDerWenseTheEuropeanPhysicalJournalA56_277}%
  \BibitemOpen
  \bibfield  {author} {\bibinfo {author} {\bibfnamefont {L.}~\bibnamefont
  {von~der Wense}}\ and\ \bibinfo {author} {\bibfnamefont {B.}~\bibnamefont
  {Seiferle}},\ }\bibfield  {title} {\bibinfo {title} {{The $ {}^{229}
  \mathrm{Th} $ isomer: prospects for a nuclear optical clock}},\ }\href
  {https://doi.org/10.1140/epja/s10050-020-00263-0} {\bibfield  {journal}
  {\bibinfo  {journal} {Eur. Phys. J. A}\ }\textbf {\bibinfo {volume} {56}},\
  \bibinfo {pages} {277} (\bibinfo {year} {2020})}\BibitemShut {NoStop}%
\bibitem [{\citenamefont {von~der Wense}\ \emph {et~al.}(2016)\citenamefont
  {von~der Wense}, \citenamefont {Seiferle}, \citenamefont {Laatiaoui},
  \citenamefont {Neumayr}, \citenamefont {Maier}, \citenamefont {Maier},
  \citenamefont {Wirth}, \citenamefont {Mokry}, \citenamefont {Runke},
  \citenamefont {Eberhardt}, \citenamefont {D\"{u}llmann}, \citenamefont
  {Trautmann},\ and\ \citenamefont {Thirolf}}]{VonDerWenseNature553_042501}%
  \BibitemOpen
  \bibfield  {author} {\bibinfo {author} {\bibfnamefont {L.}~\bibnamefont
  {von~der Wense}}, \bibinfo {author} {\bibfnamefont {B.}~\bibnamefont
  {Seiferle}}, \bibinfo {author} {\bibfnamefont {M.}~\bibnamefont {Laatiaoui}},
  \bibinfo {author} {\bibfnamefont {J.~B.}\ \bibnamefont {Neumayr}}, \bibinfo
  {author} {\bibfnamefont {H.-J.}\ \bibnamefont {Maier}}, \bibinfo {author}
  {\bibfnamefont {H.-J.}\ \bibnamefont {Maier}}, \bibinfo {author}
  {\bibfnamefont {H.-F.}\ \bibnamefont {Wirth}}, \bibinfo {author}
  {\bibfnamefont {C.}~\bibnamefont {Mokry}}, \bibinfo {author} {\bibfnamefont
  {J.}~\bibnamefont {Runke}}, \bibinfo {author} {\bibfnamefont
  {K.}~\bibnamefont {Eberhardt}}, \bibinfo {author} {\bibfnamefont {C.~E.}\
  \bibnamefont {D\"{u}llmann}}, \bibinfo {author} {\bibfnamefont {N.~G.}\
  \bibnamefont {Trautmann}},\ and\ \bibinfo {author} {\bibfnamefont {P.~G.}\
  \bibnamefont {Thirolf}},\ }\bibfield  {title} {\bibinfo {title} {{Direct
  detection of the $ {}^{229} \mathrm{Th} $ nuclear clock transition}},\ }\href
  {https://doi.org/10.1038/nature17669} {\bibfield  {journal} {\bibinfo
  {journal} {Nature}\ }\textbf {\bibinfo {volume} {553}},\ \bibinfo {pages}
  {042501} (\bibinfo {year} {2016})}\BibitemShut {NoStop}%
\bibitem [{\citenamefont {Seiferle}\ \emph {et~al.}(2017)\citenamefont
  {Seiferle}, \citenamefont {von~der Wense},\ and\ \citenamefont
  {Thirolf}}]{PhysRevLett.118.042501}%
  \BibitemOpen
  \bibfield  {author} {\bibinfo {author} {\bibfnamefont {B.}~\bibnamefont
  {Seiferle}}, \bibinfo {author} {\bibfnamefont {L.}~\bibnamefont {von~der
  Wense}},\ and\ \bibinfo {author} {\bibfnamefont {P.~G.}\ \bibnamefont
  {Thirolf}},\ }\bibfield  {title} {\bibinfo {title} {{Lifetime Measurement of
  the $^{229}\mathrm{Th}$ Nuclear Isomer}},\ }\href
  {https://doi.org/10.1103/PhysRevLett.118.042501} {\bibfield  {journal}
  {\bibinfo  {journal} {Phys. Rev. Lett.}\ }\textbf {\bibinfo {volume} {118}},\
  \bibinfo {pages} {042501} (\bibinfo {year} {2017})}\BibitemShut {NoStop}%
\bibitem [{\citenamefont {Peik}\ and\ \citenamefont
  {Tamm}(2003)}]{Peik2003Europhys.Lett.61_181}%
  \BibitemOpen
  \bibfield  {author} {\bibinfo {author} {\bibfnamefont {E.}~\bibnamefont
  {Peik}}\ and\ \bibinfo {author} {\bibfnamefont {C.}~\bibnamefont {Tamm}},\
  }\bibfield  {title} {\bibinfo {title} {{Nuclear laser spectroscopy of the 3.5
  eV transition in Th-229}},\ }\href
  {https://doi.org/10.1209/epl/i2003-00210-x} {\bibfield  {journal} {\bibinfo
  {journal} {Europhys. Lett.}\ }\textbf {\bibinfo {volume} {61}},\ \bibinfo
  {pages} {181} (\bibinfo {year} {2003})}\BibitemShut {NoStop}%
\bibitem [{\citenamefont {Dzuba}\ and\ \citenamefont
  {Flambaum}(2025{\natexlab{b}})}]{PhysRevA.111.053109}%
  \BibitemOpen
  \bibfield  {author} {\bibinfo {author} {\bibfnamefont {V.~A.}\ \bibnamefont
  {Dzuba}}\ and\ \bibinfo {author} {\bibfnamefont {V.~V.}\ \bibnamefont
  {Flambaum}},\ }\bibfield  {title} {\bibinfo {title} {{Using the Th iii ion
  for a nuclear clock and searches for new physics}},\ }\href
  {https://doi.org/10.1103/PhysRevA.111.053109} {\bibfield  {journal} {\bibinfo
   {journal} {Phys. Rev. A}\ }\textbf {\bibinfo {volume} {111}},\ \bibinfo
  {pages} {053109} (\bibinfo {year} {2025}{\natexlab{b}})}\BibitemShut
  {NoStop}%
\bibitem [{\citenamefont {Peik}\ \emph {et~al.}(2021)\citenamefont {Peik},
  \citenamefont {Schumm}, \citenamefont {Safronova}, \citenamefont
  {P\'{a}lffy}, \citenamefont {Weitenberg},\ and\ \citenamefont
  {Thirolf}}]{Peik_2021}%
  \BibitemOpen
  \bibfield  {author} {\bibinfo {author} {\bibfnamefont {E.}~\bibnamefont
  {Peik}}, \bibinfo {author} {\bibfnamefont {T.}~\bibnamefont {Schumm}},
  \bibinfo {author} {\bibfnamefont {M.~S.}\ \bibnamefont {Safronova}}, \bibinfo
  {author} {\bibfnamefont {A.}~\bibnamefont {P\'{a}lffy}}, \bibinfo {author}
  {\bibfnamefont {J.}~\bibnamefont {Weitenberg}},\ and\ \bibinfo {author}
  {\bibfnamefont {P.~G.}\ \bibnamefont {Thirolf}},\ }\bibfield  {title}
  {\bibinfo {title} {{Nuclear clocks for testing fundamental physics}},\ }\href
  {https://doi.org/10.1088/2058-9565/abe9c2} {\bibfield  {journal} {\bibinfo
  {journal} {Quantum Sci. Technol.}\ }\textbf {\bibinfo {volume} {6}},\
  \bibinfo {pages} {034002} (\bibinfo {year} {2021})}\BibitemShut {NoStop}%
\bibitem [{\citenamefont
  {Flambaum}(2006)}]{Flambaum2006PhysicalReviewLetters.97_092502}%
  \BibitemOpen
  \bibfield  {author} {\bibinfo {author} {\bibfnamefont {V.~V.}\ \bibnamefont
  {Flambaum}},\ }\bibfield  {title} {\bibinfo {title} {{Enhanced Effect of
  Temporal Variation of the Fine Structure Constant and the Strong Interaction
  in Th 229}},\ }\href {https://doi.org/10.1103/PhysRevLett.97.092502}
  {\bibfield  {journal} {\bibinfo  {journal} {Phys. Rev. Lett.}\ }\textbf
  {\bibinfo {volume} {97}},\ \bibinfo {pages} {092502} (\bibinfo {year}
  {2006})}\BibitemShut {NoStop}%
\bibitem [{\citenamefont {Tkalya}(2000)}]{TkalyaJetpLett.71_311-313}%
  \BibitemOpen
  \bibfield  {author} {\bibinfo {author} {\bibfnamefont {E.~V.}\ \bibnamefont
  {Tkalya}},\ }\bibfield  {title} {\bibinfo {title} {{Spontaneous emission
  probability for M1 transition in a dielectric medium: $^{229}\mathrm{Th}$ ($
  3/2^{+} $, $ 3.5 \pm 1.0 \, \mathrm{eV} $) decay}},\ }\href
  {https://doi.org/10.1134/1.568349} {\bibfield  {journal} {\bibinfo  {journal}
  {JETP Lett.}\ }\textbf {\bibinfo {volume} {71}},\ \bibinfo {pages} {311}
  (\bibinfo {year} {2000})}\BibitemShut {NoStop}%
\bibitem [{\citenamefont {M\"{u}ller}\ \emph {et~al.}(2017)\citenamefont
  {M\"{u}ller}, \citenamefont {Volotka}, \citenamefont {Fritzsche},\ and\
  \citenamefont {Surzhykov}}]{MULLER201784}%
  \BibitemOpen
  \bibfield  {author} {\bibinfo {author} {\bibfnamefont {R.~A.}\ \bibnamefont
  {M\"{u}ller}}, \bibinfo {author} {\bibfnamefont {A.~V.}\ \bibnamefont
  {Volotka}}, \bibinfo {author} {\bibfnamefont {S.}~\bibnamefont {Fritzsche}},\
  and\ \bibinfo {author} {\bibfnamefont {A.}~\bibnamefont {Surzhykov}},\
  }\bibfield  {title} {\bibinfo {title} {{Theoretical analysis of the electron
  bridge process in $ {}^{229} \mathrm{Th}^{3+} $}},\ }\href
  {https://doi.org/https://doi.org/10.1016/j.nimb.2017.05.004} {\bibfield
  {journal} {\bibinfo  {journal} {Nucl. Instrum. Methods Phys. Res. B}\
  }\textbf {\bibinfo {volume} {408}},\ \bibinfo {pages} {84} (\bibinfo {year}
  {2017})}\BibitemShut {NoStop}%
\bibitem [{\citenamefont {Porsev}\ \emph {et~al.}(2021)\citenamefont {Porsev},
  \citenamefont {Cheung},\ and\ \citenamefont {Safronova}}]{PORSEV2021}%
  \BibitemOpen
  \bibfield  {author} {\bibinfo {author} {\bibfnamefont {S.~G.}\ \bibnamefont
  {Porsev}}, \bibinfo {author} {\bibfnamefont {C.}~\bibnamefont {Cheung}},\
  and\ \bibinfo {author} {\bibfnamefont {M.~S.}\ \bibnamefont {Safronova}},\
  }\bibfield  {title} {\bibinfo {title} {{Low-lying energy levels of $ {}^{229}
  \mathrm{Th}^{35+} $ and the electronic bridge process}},\ }\href
  {https://doi.org/10.1088/2058-9565/ac08f1} {\bibfield  {journal} {\bibinfo
  {journal} {Quantum Sci. Technol.}\ }\textbf {\bibinfo {volume} {6}},\
  \bibinfo {pages} {034014} (\bibinfo {year} {2021})}\BibitemShut {NoStop}%
\bibitem [{\citenamefont {Porsev}\ \emph {et~al.}(2010)\citenamefont {Porsev},
  \citenamefont {Flambaum}, \citenamefont {Peik},\ and\ \citenamefont
  {Tamm}}]{PhysRevLett.105.182501}%
  \BibitemOpen
  \bibfield  {author} {\bibinfo {author} {\bibfnamefont {S.~G.}\ \bibnamefont
  {Porsev}}, \bibinfo {author} {\bibfnamefont {V.~V.}\ \bibnamefont
  {Flambaum}}, \bibinfo {author} {\bibfnamefont {E.}~\bibnamefont {Peik}},\
  and\ \bibinfo {author} {\bibfnamefont {C.}~\bibnamefont {Tamm}},\ }\bibfield
  {title} {\bibinfo {title} {{Excitation of the Isomeric $^{229m}\mathrm{Th}$
  Nuclear State via an Electronic Bridge Process in $^{229}\mathrm{Th}^{+}$}},\
  }\href {https://doi.org/10.1103/PhysRevLett.105.182501} {\bibfield  {journal}
  {\bibinfo  {journal} {Phys. Rev. Lett.}\ }\textbf {\bibinfo {volume} {105}},\
  \bibinfo {pages} {182501} (\bibinfo {year} {2010})}\BibitemShut {NoStop}%
\bibitem [{\citenamefont {Morgan}\ \emph {et~al.}(2025)\citenamefont {Morgan},
  \citenamefont {Tran~Tan}, \citenamefont {Elwell}, \citenamefont
  {Alexandrova}, \citenamefont {Hudson},\ and\ \citenamefont
  {Derevianko}}]{9s8f-hv1f}%
  \BibitemOpen
  \bibfield  {author} {\bibinfo {author} {\bibfnamefont {H.~W.~T.}\
  \bibnamefont {Morgan}}, \bibinfo {author} {\bibfnamefont {H.~B.}\
  \bibnamefont {Tran~Tan}}, \bibinfo {author} {\bibfnamefont {R.}~\bibnamefont
  {Elwell}}, \bibinfo {author} {\bibfnamefont {A.~N.}\ \bibnamefont
  {Alexandrova}}, \bibinfo {author} {\bibfnamefont {E.~R.}\ \bibnamefont
  {Hudson}},\ and\ \bibinfo {author} {\bibfnamefont {A.}~\bibnamefont
  {Derevianko}},\ }\bibfield  {title} {\bibinfo {title} {{Theory of Internal
  Conversion of the $^{229}\mathrm{Th}$ Nuclear Isomer in Solid-State Hosts}},\
  }\href {https://doi.org/10.1103/9s8f-hv1f} {\bibfield  {journal} {\bibinfo
  {journal} {Phys. Rev. Lett.}\ }\textbf {\bibinfo {volume} {134}},\ \bibinfo
  {pages} {253801} (\bibinfo {year} {2025})}\BibitemShut {NoStop}%
\bibitem [{\citenamefont {Persson}(2014{\natexlab{a}})}]{osti_1201632}%
  \BibitemOpen
  \bibfield  {author} {\bibinfo {author} {\bibfnamefont {K.}~\bibnamefont
  {Persson}},\ }\href {https://doi.org/10.17188/1201632} {\bibinfo {title}
  {{Materials Data on $ \mathrm{Ca} \mathrm{F}_2 $ (SG:225) by Materials
  Project}}} (\bibinfo {year} {2014}{\natexlab{a}})\BibitemShut {NoStop}%
\bibitem [{\citenamefont {Persson}(2014{\natexlab{b}})}]{osti_1192516}%
  \BibitemOpen
  \bibfield  {author} {\bibinfo {author} {\bibfnamefont {K.}~\bibnamefont
  {Persson}},\ }\href {https://doi.org/10.17188/1192516} {\bibinfo {title}
  {{Materials Data on $ \mathrm{Mg} \mathrm{F}_2 $ (SG:205) by Materials
  Project}}} (\bibinfo {year} {2014}{\natexlab{b}})\BibitemShut {NoStop}%
\bibitem [{\citenamefont {Persson}(2014{\natexlab{c}})}]{osti_1281377}%
  \BibitemOpen
  \bibfield  {author} {\bibinfo {author} {\bibfnamefont {K.}~\bibnamefont
  {Persson}},\ }\href {https://doi.org/10.17188/1281377} {\bibinfo {title}
  {{Materials Data on $ \mathrm{Sr} \mathrm{Li} \mathrm{Al} \mathrm{F}_6 $
  (SG:163) by Materials Project}}} (\bibinfo {year}
  {2014}{\natexlab{c}})\BibitemShut {NoStop}%
\bibitem [{\citenamefont {Hohenberg}\ and\ \citenamefont
  {Kohn}(1964)}]{Hohenberg1964Phys.Rev.136_B864}%
  \BibitemOpen
  \bibfield  {author} {\bibinfo {author} {\bibfnamefont {P.}~\bibnamefont
  {Hohenberg}}\ and\ \bibinfo {author} {\bibfnamefont {W.}~\bibnamefont
  {Kohn}},\ }\bibfield  {title} {\bibinfo {title} {{Inhomogeneous Electron
  Gas}},\ }\href {https://doi.org/10.1103/PhysRev.136.B864} {\bibfield
  {journal} {\bibinfo  {journal} {Phys. Rev.}\ }\textbf {\bibinfo {volume}
  {136}},\ \bibinfo {pages} {B864} (\bibinfo {year} {1964})}\BibitemShut
  {NoStop}%
\bibitem [{\citenamefont {Kohn}\ and\ \citenamefont
  {Sham}(1965)}]{Kohn1965Phys.Rev.140_A1133}%
  \BibitemOpen
  \bibfield  {author} {\bibinfo {author} {\bibfnamefont {W.}~\bibnamefont
  {Kohn}}\ and\ \bibinfo {author} {\bibfnamefont {L.~J.}\ \bibnamefont
  {Sham}},\ }\bibfield  {title} {\bibinfo {title} {{Self-Consistent Equations
  Including Exchange and Correlation Effects}},\ }\href
  {https://doi.org/10.1103/PhysRev.140.A1133} {\bibfield  {journal} {\bibinfo
  {journal} {Phys. Rev.}\ }\textbf {\bibinfo {volume} {140}},\ \bibinfo {pages}
  {A1133} (\bibinfo {year} {1965})}\BibitemShut {NoStop}%
\bibitem [{\citenamefont {Kohn}(1999)}]{Kohn1999Rev.Mod.Phys.71_1253}%
  \BibitemOpen
  \bibfield  {author} {\bibinfo {author} {\bibfnamefont {W.}~\bibnamefont
  {Kohn}},\ }\bibfield  {title} {\bibinfo {title} {{Nobel Lecture: Electronic
  structure of matter---wave functions and density functionals}},\ }\href
  {https://doi.org/10.1103/RevModPhys.71.1253} {\bibfield  {journal} {\bibinfo
  {journal} {Rev. Mod. Phys.}\ }\textbf {\bibinfo {volume} {71}},\ \bibinfo
  {pages} {1253} (\bibinfo {year} {1999})}\BibitemShut {NoStop}%
\bibitem [{\citenamefont {Giannozzi}\ \emph {et~al.}(2009)\citenamefont
  {Giannozzi}, \citenamefont {Baroni}, \citenamefont {Bonini}, \citenamefont
  {Calandra}, \citenamefont {Car}, \citenamefont {Cavazzoni}, \citenamefont
  {Ceresoli}, \citenamefont {Chiarotti}, \citenamefont {Cococcioni},
  \citenamefont {Dabo}, \citenamefont {Dal~Corso}, \citenamefont
  {de~Gironcoli}, \citenamefont {Fabris}, \citenamefont {Fratesi},
  \citenamefont {Gebauer}, \citenamefont {Gerstmann}, \citenamefont
  {Gougoussis}, \citenamefont {Kokalj}, \citenamefont {Lazzeri}, \citenamefont
  {Martin-Samos}, \citenamefont {Marzari}, \citenamefont {Mauri}, \citenamefont
  {Mazzarello}, \citenamefont {Paolini}, \citenamefont {Pasquarello},
  \citenamefont {Paulatto}, \citenamefont {Sbraccia}, \citenamefont {Scandolo},
  \citenamefont {Sclauzero}, \citenamefont {Seitsonen}, \citenamefont
  {Smogunov}, \citenamefont {Umari},\ and\ \citenamefont
  {Wentzcovitch}}]{Giannozzi_2009}%
  \BibitemOpen
  \bibfield  {author} {\bibinfo {author} {\bibfnamefont {P.}~\bibnamefont
  {Giannozzi}}, \bibinfo {author} {\bibfnamefont {S.}~\bibnamefont {Baroni}},
  \bibinfo {author} {\bibfnamefont {N.}~\bibnamefont {Bonini}}, \bibinfo
  {author} {\bibfnamefont {M.}~\bibnamefont {Calandra}}, \bibinfo {author}
  {\bibfnamefont {R.}~\bibnamefont {Car}}, \bibinfo {author} {\bibfnamefont
  {C.}~\bibnamefont {Cavazzoni}}, \bibinfo {author} {\bibfnamefont
  {D.}~\bibnamefont {Ceresoli}}, \bibinfo {author} {\bibfnamefont {G.~L.}\
  \bibnamefont {Chiarotti}}, \bibinfo {author} {\bibfnamefont {M.}~\bibnamefont
  {Cococcioni}}, \bibinfo {author} {\bibfnamefont {I.}~\bibnamefont {Dabo}},
  \bibinfo {author} {\bibfnamefont {A.}~\bibnamefont {Dal~Corso}}, \bibinfo
  {author} {\bibfnamefont {S.}~\bibnamefont {de~Gironcoli}}, \bibinfo {author}
  {\bibfnamefont {S.}~\bibnamefont {Fabris}}, \bibinfo {author} {\bibfnamefont
  {G.}~\bibnamefont {Fratesi}}, \bibinfo {author} {\bibfnamefont
  {R.}~\bibnamefont {Gebauer}}, \bibinfo {author} {\bibfnamefont
  {U.}~\bibnamefont {Gerstmann}}, \bibinfo {author} {\bibfnamefont
  {C.}~\bibnamefont {Gougoussis}}, \bibinfo {author} {\bibfnamefont
  {A.}~\bibnamefont {Kokalj}}, \bibinfo {author} {\bibfnamefont
  {M.}~\bibnamefont {Lazzeri}}, \bibinfo {author} {\bibfnamefont
  {L.}~\bibnamefont {Martin-Samos}}, \bibinfo {author} {\bibfnamefont
  {N.}~\bibnamefont {Marzari}}, \bibinfo {author} {\bibfnamefont
  {F.}~\bibnamefont {Mauri}}, \bibinfo {author} {\bibfnamefont
  {R.}~\bibnamefont {Mazzarello}}, \bibinfo {author} {\bibfnamefont
  {S.}~\bibnamefont {Paolini}}, \bibinfo {author} {\bibfnamefont
  {A.}~\bibnamefont {Pasquarello}}, \bibinfo {author} {\bibfnamefont
  {L.}~\bibnamefont {Paulatto}}, \bibinfo {author} {\bibfnamefont
  {C.}~\bibnamefont {Sbraccia}}, \bibinfo {author} {\bibfnamefont
  {S.}~\bibnamefont {Scandolo}}, \bibinfo {author} {\bibfnamefont
  {G.}~\bibnamefont {Sclauzero}}, \bibinfo {author} {\bibfnamefont {A.~P.}\
  \bibnamefont {Seitsonen}}, \bibinfo {author} {\bibfnamefont {A.}~\bibnamefont
  {Smogunov}}, \bibinfo {author} {\bibfnamefont {P.}~\bibnamefont {Umari}},\
  and\ \bibinfo {author} {\bibfnamefont {R.~M.}\ \bibnamefont {Wentzcovitch}},\
  }\bibfield  {title} {\bibinfo {title} {{QUANTUM ESPRESSO: a modular and
  open-source software project for quantum simulations of materials}},\ }\href
  {https://doi.org/10.1088/0953-8984/21/39/395502} {\bibfield  {journal}
  {\bibinfo  {journal} {J. Phys.: Condens. Matter}\ }\textbf {\bibinfo {volume}
  {21}},\ \bibinfo {pages} {395502} (\bibinfo {year} {2009})}\BibitemShut
  {NoStop}%
\bibitem [{\citenamefont {Perdew}\ \emph
  {et~al.}(1996{\natexlab{a}})\citenamefont {Perdew}, \citenamefont {Burke},\
  and\ \citenamefont {Wang}}]{Perdew1996Phys.Rev.B54_16533}%
  \BibitemOpen
  \bibfield  {author} {\bibinfo {author} {\bibfnamefont {J.~P.}\ \bibnamefont
  {Perdew}}, \bibinfo {author} {\bibfnamefont {K.}~\bibnamefont {Burke}},\ and\
  \bibinfo {author} {\bibfnamefont {Y.}~\bibnamefont {Wang}},\ }\bibfield
  {title} {\bibinfo {title} {{Generalized gradient approximation for the
  exchange-correlation hole of a many-electron system}},\ }\href
  {https://doi.org/10.1103/PhysRevB.54.16533} {\bibfield  {journal} {\bibinfo
  {journal} {Phys. Rev. B}\ }\textbf {\bibinfo {volume} {54}},\ \bibinfo
  {pages} {16533} (\bibinfo {year} {1996}{\natexlab{a}})}\BibitemShut {NoStop}%
\bibitem [{\citenamefont {Perdew}\ \emph
  {et~al.}(1996{\natexlab{b}})\citenamefont {Perdew}, \citenamefont {Burke},\
  and\ \citenamefont {Ernzerhof}}]{Perdew1996Phys.Rev.Lett.77_3865}%
  \BibitemOpen
  \bibfield  {author} {\bibinfo {author} {\bibfnamefont {J.~P.}\ \bibnamefont
  {Perdew}}, \bibinfo {author} {\bibfnamefont {K.}~\bibnamefont {Burke}},\ and\
  \bibinfo {author} {\bibfnamefont {M.}~\bibnamefont {Ernzerhof}},\ }\bibfield
  {title} {\bibinfo {title} {{Generalized Gradient Approximation Made
  Simple}},\ }\href {https://doi.org/10.1103/PhysRevLett.77.3865} {\bibfield
  {journal} {\bibinfo  {journal} {Phys. Rev. Lett.}\ }\textbf {\bibinfo
  {volume} {77}},\ \bibinfo {pages} {3865} (\bibinfo {year}
  {1996}{\natexlab{b}})}\BibitemShut {NoStop}%
\bibitem [{\citenamefont {Torrent}\ \emph {et~al.}(2010)\citenamefont
  {Torrent}, \citenamefont {Holzwarth}, \citenamefont {Jollet}, \citenamefont
  {Harris}, \citenamefont {Lepley},\ and\ \citenamefont
  {Xu}}]{TORRENT20101862}%
  \BibitemOpen
  \bibfield  {author} {\bibinfo {author} {\bibfnamefont {M.}~\bibnamefont
  {Torrent}}, \bibinfo {author} {\bibfnamefont {N.~A.~W.}\ \bibnamefont
  {Holzwarth}}, \bibinfo {author} {\bibfnamefont {F.}~\bibnamefont {Jollet}},
  \bibinfo {author} {\bibfnamefont {D.}~\bibnamefont {Harris}}, \bibinfo
  {author} {\bibfnamefont {N.}~\bibnamefont {Lepley}},\ and\ \bibinfo {author}
  {\bibfnamefont {X.}~\bibnamefont {Xu}},\ }\bibfield  {title} {\bibinfo
  {title} {{Electronic structure packages: Two implementations of the projector
  augmented wave (PAW) formalism}},\ }\href
  {https://doi.org/10.1016/j.cpc.2010.07.036} {\bibfield  {journal} {\bibinfo
  {journal} {Comput. Phys. Commun.}\ }\textbf {\bibinfo {volume} {181}},\
  \bibinfo {pages} {1862} (\bibinfo {year} {2010})}\BibitemShut {NoStop}%
\bibitem [{\citenamefont
  {Dal~Corso}(2014)}]{DalCorso2014Comput.Mater.Sci.95_337}%
  \BibitemOpen
  \bibfield  {author} {\bibinfo {author} {\bibfnamefont {A.}~\bibnamefont
  {Dal~Corso}},\ }\bibfield  {title} {\bibinfo {title} {{Pseudopotentials
  periodic table: From H to Pu}},\ }\href
  {https://doi.org/10.1016/j.commatsci.2014.07.043} {\bibfield  {journal}
  {\bibinfo  {journal} {Comput. Mater. Sci.}\ }\textbf {\bibinfo {volume}
  {95}},\ \bibinfo {pages} {337} (\bibinfo {year} {2014})}\BibitemShut
  {NoStop}%
\bibitem [{\citenamefont {te~Velde}\ \emph {et~al.}(2001)\citenamefont
  {te~Velde}, \citenamefont {Bickelhaupt}, \citenamefont {Baerends},
  \citenamefont {Fonseca~Guerra}, \citenamefont {van Gisbergen}, \citenamefont
  {Snijders},\ and\ \citenamefont
  {Ziegler}}]{https://doi.org/10.1002/jcc.1056}%
  \BibitemOpen
  \bibfield  {author} {\bibinfo {author} {\bibfnamefont {G.}~\bibnamefont
  {te~Velde}}, \bibinfo {author} {\bibfnamefont {F.~M.}\ \bibnamefont
  {Bickelhaupt}}, \bibinfo {author} {\bibfnamefont {E.~J.}\ \bibnamefont
  {Baerends}}, \bibinfo {author} {\bibfnamefont {C.}~\bibnamefont
  {Fonseca~Guerra}}, \bibinfo {author} {\bibfnamefont {S.~J.~A.}\ \bibnamefont
  {van Gisbergen}}, \bibinfo {author} {\bibfnamefont {J.~G.}\ \bibnamefont
  {Snijders}},\ and\ \bibinfo {author} {\bibfnamefont {T.}~\bibnamefont
  {Ziegler}},\ }\bibfield  {title} {\bibinfo {title} {{Chemistry with ADF}},\
  }\href {https://doi.org/10.1002/jcc.1056} {\bibfield  {journal} {\bibinfo
  {journal} {J. Comput. Chem.}\ }\textbf {\bibinfo {volume} {22}},\ \bibinfo
  {pages} {931} (\bibinfo {year} {2001})}\BibitemShut {NoStop}%
\bibitem [{\citenamefont {Van~Lenthe}\ and\ \citenamefont
  {Baerends}(2003)}]{https://doi.org/10.1002/jcc.10255}%
  \BibitemOpen
  \bibfield  {author} {\bibinfo {author} {\bibfnamefont {E.}~\bibnamefont
  {Van~Lenthe}}\ and\ \bibinfo {author} {\bibfnamefont {E.~J.}\ \bibnamefont
  {Baerends}},\ }\bibfield  {title} {\bibinfo {title} {{Optimized Slater-type
  basis sets for the elements 1--118}},\ }\href
  {https://doi.org/10.1002/jcc.10255} {\bibfield  {journal} {\bibinfo
  {journal} {J. Comput. Chem.}\ }\textbf {\bibinfo {volume} {24}},\ \bibinfo
  {pages} {1142} (\bibinfo {year} {2003})}\BibitemShut {NoStop}%
\bibitem [{\citenamefont {Becke}(1993)}]{10.1063/1.464913}%
  \BibitemOpen
  \bibfield  {author} {\bibinfo {author} {\bibfnamefont {A.~D.}\ \bibnamefont
  {Becke}},\ }\bibfield  {title} {\bibinfo {title} {{Density-functional
  thermochemistry. III. The role of exact exchange}},\ }\href
  {https://doi.org/10.1063/1.464913} {\bibfield  {journal} {\bibinfo  {journal}
  {J. Chem. Phys.}\ }\textbf {\bibinfo {volume} {98}},\ \bibinfo {pages} {5648}
  (\bibinfo {year} {1993})}\BibitemShut {NoStop}%
\bibitem [{\citenamefont {van Lenthe}\ \emph {et~al.}(1993)\citenamefont {van
  Lenthe}, \citenamefont {Baerends},\ and\ \citenamefont
  {Snijders}}]{10.1063/1.466059}%
  \BibitemOpen
  \bibfield  {author} {\bibinfo {author} {\bibfnamefont {E.}~\bibnamefont {van
  Lenthe}}, \bibinfo {author} {\bibfnamefont {E.~J.}\ \bibnamefont
  {Baerends}},\ and\ \bibinfo {author} {\bibfnamefont {J.~G.}\ \bibnamefont
  {Snijders}},\ }\bibfield  {title} {\bibinfo {title} {{Relativistic regular
  two-component Hamiltonians}},\ }\href {https://doi.org/10.1063/1.466059}
  {\bibfield  {journal} {\bibinfo  {journal} {J. Chem. Phys.}\ }\textbf
  {\bibinfo {volume} {99}},\ \bibinfo {pages} {4597} (\bibinfo {year}
  {1993})}\BibitemShut {NoStop}%
\bibitem [{\citenamefont {Rodr\'{\i}guez}\ \emph {et~al.}(2009)\citenamefont
  {Rodr\'{\i}guez}, \citenamefont {Bader}, \citenamefont {Ayers}, \citenamefont
  {Michel}, \citenamefont {G\"{o}tz},\ and\ \citenamefont
  {Bo}}]{RODRIGUEZ2009149}%
  \BibitemOpen
  \bibfield  {author} {\bibinfo {author} {\bibfnamefont {J.~I.}\ \bibnamefont
  {Rodr\'{\i}guez}}, \bibinfo {author} {\bibfnamefont {R.~F.}\ \bibnamefont
  {Bader}}, \bibinfo {author} {\bibfnamefont {P.~W.}\ \bibnamefont {Ayers}},
  \bibinfo {author} {\bibfnamefont {C.}~\bibnamefont {Michel}}, \bibinfo
  {author} {\bibfnamefont {A.~W.}\ \bibnamefont {G\"{o}tz}},\ and\ \bibinfo
  {author} {\bibfnamefont {C.}~\bibnamefont {Bo}},\ }\bibfield  {title}
  {\bibinfo {title} {{A high performance grid-based algorithm for computing
  QTAIM properties}},\ }\href {https://doi.org/10.1016/j.cplett.2009.02.081}
  {\bibfield  {journal} {\bibinfo  {journal} {Chem. Phys. Lett.}\ }\textbf
  {\bibinfo {volume} {472}},\ \bibinfo {pages} {149} (\bibinfo {year}
  {2009})}\BibitemShut {NoStop}%
\bibitem [{\citenamefont {Dessovic}\ \emph {et~al.}(2014)\citenamefont
  {Dessovic}, \citenamefont {Mohn}, \citenamefont {Jackson}, \citenamefont
  {Winkler}, \citenamefont {Schreitl}, \citenamefont {Kazakov},\ and\
  \citenamefont {Schumm}}]{Dessovic_2014}%
  \BibitemOpen
  \bibfield  {author} {\bibinfo {author} {\bibfnamefont {P.}~\bibnamefont
  {Dessovic}}, \bibinfo {author} {\bibfnamefont {P.}~\bibnamefont {Mohn}},
  \bibinfo {author} {\bibfnamefont {R.~A.}\ \bibnamefont {Jackson}}, \bibinfo
  {author} {\bibfnamefont {G.}~\bibnamefont {Winkler}}, \bibinfo {author}
  {\bibfnamefont {M.}~\bibnamefont {Schreitl}}, \bibinfo {author}
  {\bibfnamefont {G.}~\bibnamefont {Kazakov}},\ and\ \bibinfo {author}
  {\bibfnamefont {T.}~\bibnamefont {Schumm}},\ }\bibfield  {title} {\bibinfo
  {title} {{229Thorium-doped calcium fluoride for nuclear laser
  spectroscopy}},\ }\href {https://doi.org/10.1088/0953-8984/26/10/105402}
  {\bibfield  {journal} {\bibinfo  {journal} {J. Phys.: Condens. Matter}\
  }\textbf {\bibinfo {volume} {26}},\ \bibinfo {pages} {105402} (\bibinfo
  {year} {2014})}\BibitemShut {NoStop}%
\bibitem [{\citenamefont {Momma}\ and\ \citenamefont
  {Izumi}(2011)}]{Momma:db5098}%
  \BibitemOpen
  \bibfield  {author} {\bibinfo {author} {\bibfnamefont {K.}~\bibnamefont
  {Momma}}\ and\ \bibinfo {author} {\bibfnamefont {F.}~\bibnamefont {Izumi}},\
  }\bibfield  {title} {\bibinfo {title} {{VESTA3 for three-dimensional
  visualization of crystal, volumetric and morphology data}},\ }\href
  {https://doi.org/10.1107/S0021889811038970} {\bibfield  {journal} {\bibinfo
  {journal} {J. Appl. Crystallogr.}\ }\textbf {\bibinfo {volume} {44}},\
  \bibinfo {pages} {1272} (\bibinfo {year} {2011})}\BibitemShut {NoStop}%
\end{thebibliography}
\end{document}